\pdfoutput=1
\documentclass[a4paper, 11pt, oneside]{Thesis}  
\usepackage{listings}
\usepackage{color}
\usepackage[export]{adjustbox}
\usepackage{tikz}
\usepackage{hologo}
\usepackage{graphicx}

\usetikzlibrary{positioning}
\definecolor{dkgreen}{rgb}{0,0.6,0}
\definecolor{gray}{rgb}{0.5,0.5,0.5}
\definecolor{mauve}{rgb}{0.58,0,0.82}

\usepackage[square, numbers, comma, sort&compress]{natbib}  

\usepackage{fancyhdr}
\usepackage{verbatim}  
\usepackage{vector}  
\title{Citation Data-set for Machine Learning Citation Styles and Entity Extraction from Citation Strings}
\author{Niall Martin Ryan}
\date{April 2018}

\begin{document}

\begin{titlepage}
    \begin{center}
        \LARGE
        \textbf{}
        
        \vspace{2cm}
        
        \includegraphics[width=1.0\textwidth]{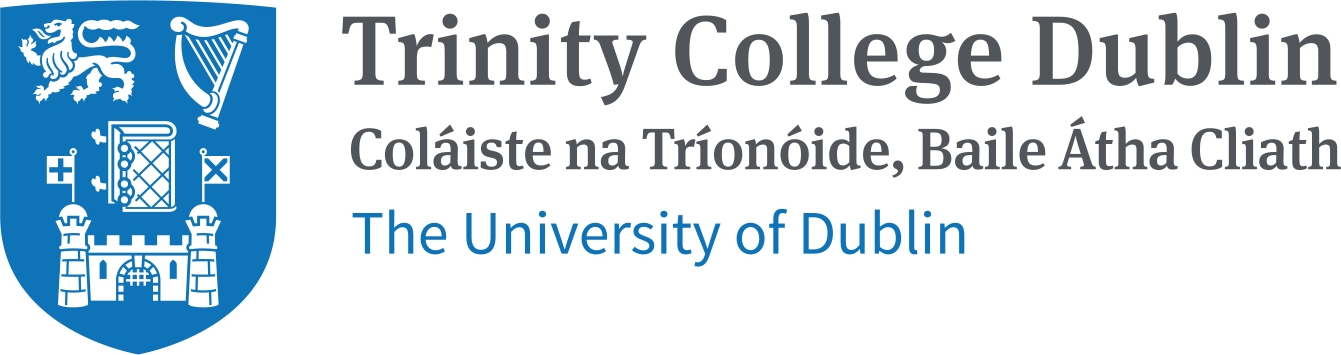}
        \vspace{0.5cm}

        \huge

        \vspace{2cm}
        \LARGE
        \textbf{Citation Data-set for Machine Learning Citation Styles and Entity Extraction from Citation Strings\\}
        \vspace{1cm}
        \Large
        Niall Martin Ryan\\
        B.A.(Mod.) Computer Science\\
        Final Year Project  April 2018\\
        Supervisor: Prof Dr Joeran Beel\\
        Co-Supervisor: Dr Dominika Tkaczyk
        \vfill
        \vspace{1cm}
        \large
        \textbf{
        School of Computer Science and Statistics\\
        O’Reilly Institute, Trinity College, Dublin 2, Ireland\\
        }
        
    \end{center}
\end{titlepage}

\setstretch{1.3}  

\fancyhead{}  
\rhead{\thepage}  
\lhead{}  

\pagestyle{fancy}  

\Declaration{

\addtocontents{toc}{\vspace{1em}}  

I hereby declare that this project is entirely my own work and that it has not been submitted
as an exercise for a degree at this or any other university 
\\

Signed:\\
\rule[1em]{25em}{0.5pt}  
 
Date:\\
\rule[1em]{25em}{0.5pt}  
\\Niall Martin Ryan
}

\clearpage  


\addtotoc{Abstract}  
\abstract{
\addtocontents{toc}{\vspace{1em}}  

Citation parsing is fundamental for search engines within academia and the protection of intellectual property. Meticulous extraction is further needed when evaluating the similarity of documents and calculating their citation impact. Citation parsing involves the identification and dissection of citation strings into their bibliography components such as "Author", "Volume","Title", etc. This meta-data can be difficult to acquire accurately due to the thousands of different styles and noise that can be applied to a bibliography to create the citation string. Many approaches exist already to accomplish accurate parsing of citation strings. This dissertation describes the creation of a large data-set which can be used to aid in the training of these approaches which have limited data.It also describes the investigation into if the downfall of these approaches to citation parsing and in particular the machine learning based approaches is because of the lack of size associated to the data used to train them.  

}

\clearpage  

\setstretch{1.3}  

\acknowledgements{
\addtocontents{toc}{\vspace{1em}}  

Firstly, I would like to thank my supervisor J{\"o}ran Beel. I could not have done this without his guidance, expertise, assistance and generous availability for queries and meetings alike.\par
I'd also like to thank Dominika Tkaczyk for her help throughout this project, especially for insight and experience within the field.\par
I would like to thank my friend Owen Mooney for helping me get the project off the ground with his advice about scraping. \par
Thank you to Michel Krämer, for answering my queries about the software he has developed for \hologo{BibTeX}/citation manipulation.\par
I thank Conor Fulham for his immaculate proof reading and for carrying me on tough days. \par
Finally, I would like to thank my family for their constant support and tolerance of me over the duration of this project. Especially my father Declan Ryan for his helpful planning and late night discussions.\par

}
\clearpage  

\pagestyle{fancy}  

\lhead{\emph{Contents}}  
\tableofcontents  

\lhead{\emph{List of Figures}}  
\listoffigures  

\lhead{\emph{List of Tables}}  
\listoftables  

\setstretch{1.5}  
\lhead{\emph{Abbreviations}}  
\listofsymbols{ll}  
{
\textbf{AI} & \textbf{A}rtificial \textbf{I}ntelligence \\
\textbf{CAPTCHA} & \textbf{C}ompletely \textbf{A}utomated \textbf{P}ublic \textbf{T}uring test to tell \textbf{C}omputers and \textbf{H}umans \textbf{A}part \\
\textbf{CRF} & \textbf{C}onditional \textbf{R}andom \textbf{F}ield \\
\textbf{CSL} & \textbf{C}itation \textbf{S}tyle \textbf{L}anguage \\
\textbf{DDOS} & \textbf{D}istributed \textbf{D}enial \textbf{O}f \textbf{S}ervice\\
\textbf{DOS} & \textbf{D}enial \textbf{O}f \textbf{S}ervice\\
\textbf{HMM} & \textbf{H}idden \textbf{M}arkov \textbf{M}odel\\
\textbf{IP} & \textbf{I}nternet \textbf{P}rotocol \\
\textbf{RAM} & \textbf{R}andom \textbf{A}ccess \textbf{M}emory \\
\textbf{RID} & \textbf{R}andom \textbf{I}cremental \textbf{D}elay \\
\textbf{SVM} & \textbf{S}upport \textbf{V}ector \textbf{M}achines \\
\textbf{TD} & \textbf{T}hreshold \textbf{D}elay  \\
\textbf{XML} & e\textbf{X}tensible \textbf{M}arkup \textbf{L}anguage \\


%
}


\setstretch{1.3}  

\dedicatory{I dedicate this to my family. Declan, Caroline and Aidan for their love and support}

\addtocontents{toc}{\vspace{2em}}  

\mainmatter	  
\pagestyle{fancy}  
\fancyhf{}
\fancyhead[LE,RO]{Niall Martin Ryan}
\fancyhead[RE,LO]{Citation Data-set for Machine Learning Citation Styles}
\cfoot{\thepage}

\chapter{Introduction}

In the past decade there has been an explosion in the amount of scientific publications accessible on many different library websites. In order to search and use these libraries effectively there is a amplifying demand for accurate organization. Organizing these articles by using the meta-data associated with them, allows for more efficient searching, calculations of citation impact and recommending similar articles \cite{Beel2014a}. Many of the most accurate citation parsers or meta data extraction approaches are through machine learning strategies and systems \cite{outofbox}. The overall accuracy of machine learning based approaches depends heavily on large amounts of diverse training data. In Tkaczyk \textit{et al}\cite{outofbox}, a data-set of near 400,000 unique reference strings is used to compare 'out of the box' tools vs trained versions of the same tools. Their results found increases of 2-4\% for already very accurate tools and up to 16\% for the weaker 'out of the box' tools. They are looking in subsequent research to use other larger data-sets which are lacking in this area. Cermine \cite{DBLP:journals/ijdar/TkaczykSFDB15, DBLP:conf/das/TkaczykSDFB14, DBLP:conf/esws/TkaczykB15} is an open source meta data extraction system which aims to deal with issues in this area, such as the large amount of possible layouts and styles that can be used, as well as the complications that the PDF format brings by not preserving delicate information within articles. Particularly the documents structure and layout which impede upon meta data extraction. This information usually needs to be reverse engineered for it to be recovered. They used a data-set referred to as GROTOAP2 \cite{tkaczyk2014grotoap2} which contained "13,210 ground truth files" within its sample. It also contained errors because of non manual preparation/altering which meant it had both segmentation and labelling errors. Cermine combined 3 small open source data sets for its reference parsing evaluation.CiteSeer \cite{Giles98citeseer}, Cora-ref \cite{mccallum2000automating} and open source PubMed \cite{pubmed} references. Cora-ref contained a total of 50,000 collected papers. These are not only out of date, but rather small in comparison to the data-sets that exist in other areas of machine learning. Which are truly needed in order to receive accurate results.   
ParsCit \cite{councill2008parscit} is another machine learning based citation parser which is planning to expand there successful research into larger training data-sets. ParsCit is currently implemented in a digital library called CiteSeerX. 
The difficulty of reference parsing and why this data-set is needed within this area is firstly due to the sheer number of different styles that are used within the academic community, that span in the thousands. Secondly because of human error and noise created by automated sources, a rule/template approach cannot account for these kind of situations. A machine learning based approach is better suited to learning and adapting to the discrepancies within large data-sets 

This project aims to help and improve on current meta-data extraction techniques by creating a very large reference data-set to train on machine learning based citation parsing tools. \par
\clearpage
\textbf{This shows examples of just a few stylings for the same reference}
\begin{lstlisting}
Argon C. & McLaughlin S. W. 2002. A parallel decoder for low latency decoding of turbo product codes. IEEE Communications Letters 6: 70–72.
Argon C, McLaughlin SW. 2002. A parallel decoder for low latency decoding of turbo product codes. IEEE Communications Letters. 6(2):70–72 
Argon, C. and McLaughlin, S. W. (2002), 'A parallel decoder for low latency decoding of turbo product codes', IEEE Communications Letters, 6:2, pp. 70–72. https://doi.org/10.1109/4234.984698 
1. Argon C, McLaughlin SW. A parallel decoder for low latency decoding of turbo product codes. IEEE Communications Letters 2002; 6: 70–2. Available at: https://doi.org/10.1109/4234.984698. 
Argon, Cenk, and Steven W. McLaughlin 2002 A parallel decoder for low latency decoding of turbo product codes. IEEE Communications Letters 6(2). IEEE Communications Letters: 70–72. Retrieved. from https://doi.org/10.1109/4234.984698. 
ARGON, Cenk; MCLAUGHLIN, Steven W. A parallel decoder for low latency decoding of turbo product codes. IEEE Communications Letters, IEEE Communications Letters. v. 6, n. 2, p. 70–72, 2002. Disponível em: <https://doi.org/10.1109/4234.984698>. 
1. Argon C, McLaughlin SW. A parallel decoder for low latency decoding of turbo product codes. 2002;6:70–72.  
Argon C., McLaughlin S.W., 2002. A parallel decoder for low latency decoding of turbo product codes. IEEE Communications Letters, 6 (2): 70‑72, doi: 10.1109/4234.984698 
\end{lstlisting}
\par
\begin{center}
    Figure \ref{CitationStyleRefExamples} \label{CitationStyleRefExamples} - A single reference shown in different example citation styles
    
    \ref{CitationStyleRefExamples}
\end{center}

\section{Motivation}
Accurate citation parsing is a necessity within academic search engines and for the security of intellectual property. ‘Automated extraction of bibliographic data, such as article titles, author names, abstracts, and references is essential to the affordable creation of large citation databases’ [1]. It aids in the identification of correlating documents and evaluating the citation impact of publications by researchers/scholars. For more effective research methods in this area, categorization and the highlighting of articles with strong correlation is crucial. Fedoryszak \textit{et al} \cite{DBLP:conf/ercimdl/FedoryszakTB13, DBLP:series/sci/FedoryszakBTW13} reasons that parsing is generally the first step before citing documents are identified within a collection, breaking up the process into phases. segmentation and entity resolution. In another paper they describe how parsing is pivotal to a "matchers" effectiveness. The parser attempts to identify fragments of raw citation strings which contain substantial information such as author, year, source.

“Citation parsing” refers to identifying and extracting a reference from the full text and classifying its elements into fields such as author, journal, publication year etc. from the bibliography. To offer an example of a citation parsers function, the parser may likely identify the citation marker [2] first but in regards to this area it may just begin with extraction of the bibliography, extract from the bibliography the first entry 'Galli', 'Luca' and 'Fraternali' which are the authors from \ref{fig:ExampleReference}. Figure \ref{fig:ExampleBibTeX} is an example of one such \hologo{BibTeX} sample which provides meta-data about the document. Many applications feature document retrieval elements, such as recommendation systems and search engines. 
Recommendation systems depend immensely upon the accuracy of references and reference parsing. Within Beel \textit{et al} \cite{Beel2014a, Beel2011b} they describe the need for open source massive access to citation data and full scholarly articles. It relies heavily on accurate references to resolve "dis-ambiguity of author names" as well as identification of duplicate papers. Inside Mr. DLib a open source, non-profit recommendation system as a product, describes recommender systems in academia as helping "researchers and scientists overcome information overload" \cite{Beel2017g}. Recommendation systems depend on accurate citation parsing due to the need to calculate cite proximity, making accurate connections without confusion of reference meta data as well as to maximize the quantity of documents recommended.
 \begin{figure}
 \caption{Example of Reference}\label{fig:ExampleReference}
\begin{lstlisting}
"[0] - Galli, Luca and Fraternali, Piero and Bozzon, Alessandro, “On the Application of Game Mechanics in Information Retrieval”, in Proceedings of the GamifIR 14, 2014."
\end{lstlisting}
\end{figure}
"Citations play important roles for both the rhetorical structure and the semantic content of the articles, and as such, citation information has shown to benefit many text mining tasks including information retrieval, information extraction, summarization, and question answering"\cite{zhang2011structural}.\par
\begin{figure}
\begin{center}
\caption{Example \hologo{BibTeX} Entry}\label{fig:ExampleBibTeX}
\includegraphics[scale=0.6]{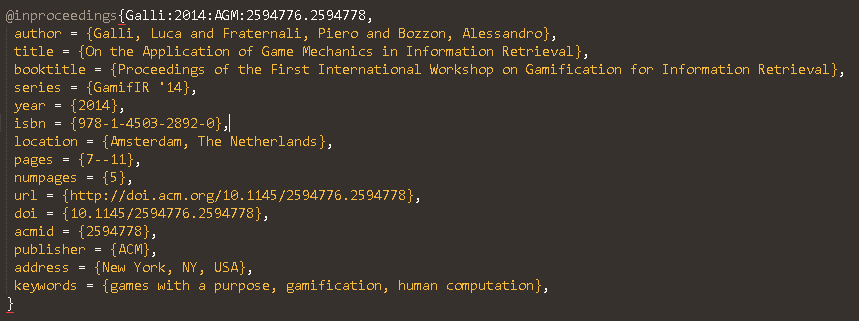}
\end{center}
\end{figure}

\section{Research Problem}
Over the years many approaches to reference parsing have been proposed, including regular expressions, knowledge-based approaches and supervised machine learning. Machine learning-based solutions, in particular those falling into the category of supervised sequence tagging, are considered a state-of-the-art technique for reference parsing. They still suffer from two issues: the lack of sufficiently big and diverse data sets and problems with generalization to unseen reference formats. In particular the need for an 'open source', 'massive' data set is the main subject of this research paper. Cora is one example of a small, outdated data set which is still used in recent years \cite{mccallum2000automating}, containing 50,000 PDFs. CiteSeerX, Wu \textit{et al} \cite{IAAI148607} currently contains over 4 million documents containing citations, at a nice moderate size but still needs to be parsed and styled in order to train an ML algorithm. An 'out-of-the-box' massive, open source training data set is crucial for the improvement of citation parsing tools based on CRF machine learning  approaches. Lipinski \textit{et al} \cite{Lipinski:2013:EHM:2467696.2467753} uses {\tt arXiv} as it's data set for training and evaluation. An extremely small set was used for evaluation testing, "we obtained 1,153 random PDF articles including their metadata, dated from 2006 to 2010". Yu \textit{et al} \cite{Yu2007} used the cora dataset as well as one by Han, They used 350 "reference items" randomly for training and 150 for evaluation testing. \cite{Zhang2011} mentions a manually built database "Institute for Scientific Information" and also that of citeSeer using 2750 citation strings as training data. Zhang created a database by automating scraping of PubMed journals, randomly selecting 2\% from each journal so that it was diverse. 672 articles were acquired which lead to a total number of citations strings of 27,606. This is rather small data set with impressive results. A last noteable comment is made about supervised machine learning algorithms being "generalizable and the learning framework is robust, as long as training data is made available" \cite{Zhang2011} which comes back to the idea that Cameron spoke about in which free open source data sets and libraries should be made available. Lin \textit{et al} \cite{Lin:2010:EFF:1867735.1867749}, obtains an initial 260 articles from PubMed which were subsequently dropped to 93 which were manually annotated and used for training and evaluation. In Pinto \textit{et al} \cite{pinto2003table} they use CRFs and refer to how they use sums which contain "gaussian prior over parameters" and variance in order to help with the scarcity of training data available. Penga \textit{et al} (2013) \cite{pengaccurate} were still using benchmark data sets that were small and out of date, compensating in other ways to make up for the lack of data, which happens to also use CRFs in their approach. This quote sums up the situation of small datasets "This is particularly important for the scientific publishing domain, where currently most of the datasets are being created in an author-driven, manual manner" \cite{groza2012reference}. They also describe the influence of the cora data-set at the time, "The CORA dataset is the first gold standard created for automatic reference chunking. It comprises two hundred reference strings and focuses on the computer science area" \cite{groza2012reference}.
\par 
Papers currently evaluating and improving upon CRF based tools are using data sets which are not publicly available \cite{outofbox}. It would appear that those data sets are being used for commercial use, which may or may not benefit the progression of technology in this area. Google Scholar must be the most prolific for easy access to individual \hologo{BibTeX} entries but remains under lock and key as could be expected.


For deep learning, much larger data-sets would be needed to train a sophisticated model than exist today. Early research in this area used rule-based applications which is loosely based around the expert knowledge of someone in a specific domain. Patterns tend to emerge, that the expert can then formulate in a method to parse such strings. Rule-based methods have some draw-backs and only really have success in specific small to moderate journal areas due to the fact that journal publishers regularly are given specifics on what predefined citation styles to use. These rule-based methods are also less adaptable and are difficult to optimize under these circumstances. 


\section{Research Goal}

\begin{figure}[h]
\begin{center}
\caption{Design Schema} \label{fig:DesignSchema}
\includegraphics[scale=.70]{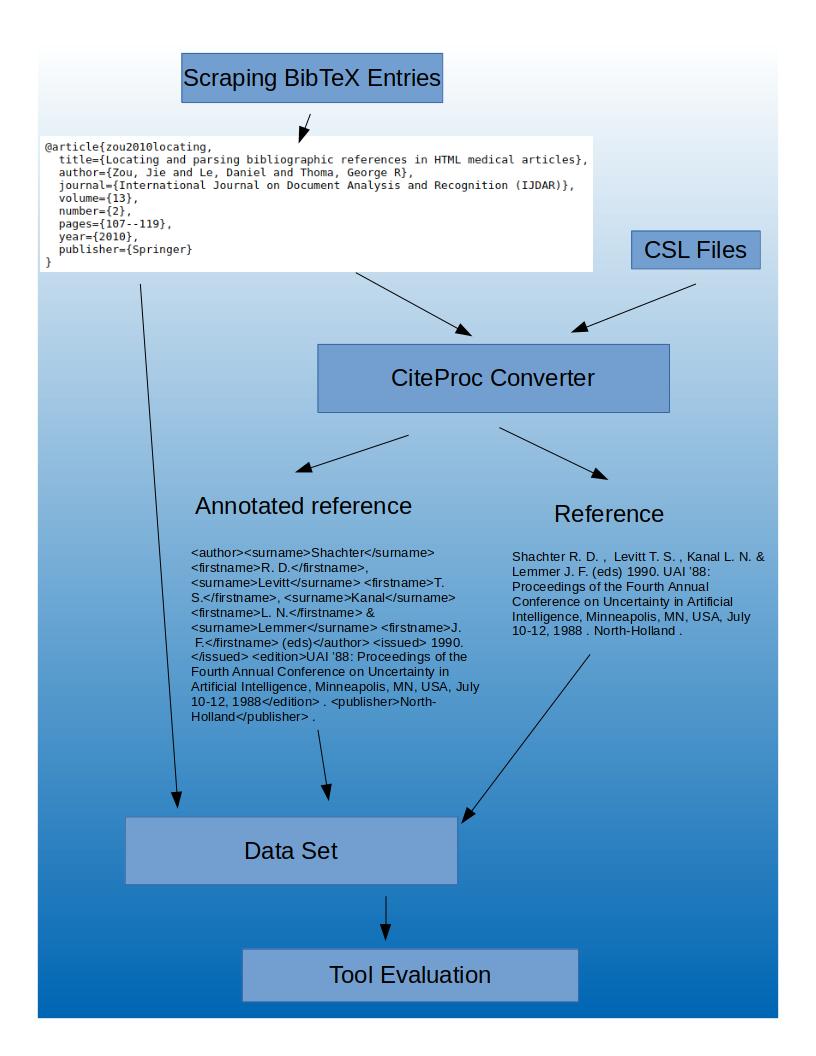}
\end{center}
\end{figure}

The research goal is to build a massive, open source data set which can be used for training high end citation parsing machine learning tools and evaluate their performance. This goal can be broken into gathering \hologo{BibTeX} entries, citation generation into thousands of different styles and annotating of citations. The \hologo{BibTeX} entries can be scraped from many different electronic libraries and recommender systems. These \hologo{BibTeX} entries are the meta data we plan to retrieve when the citation parsing tools are finished with the references. In reference to the Figure \ref{fig:DesignSchema}, Annotated and Standard references are needed in order to train the machine learning based tools. These references are obtained by ways of CiteProc \cite{michel}, which takes in a CSL template file and a \hologo{BibTeX} entry, returning the output of a reference string. Each \hologo{BibTeX} will be converted into thousands of stylized references. Then the data set can be used to evaluate citation parsing tools using the annotated and standard references.

\clearpage

Figure \ref{ProjectSchemaLayout} \label{ProjectSchemaLayout} - Project Schema layout - \hologo{BibTeX} entries are scraped from the internet. The CiteProc Converter takes in these \hologo{BibTeX} entries along with a standard CSL file for output of a reference and also a altered CSL file which will give output as an annotated reference. Both of these outputs are stored in the data-set alongside the original scraped \hologo{BibTeX}. The Data-set is then passed on for evaluation testing. 

\clearpage


\chapter{State of the Art}
Numerous methodology have been used in research articles to solve the generalized problem of meta data extraction in reference strings. One of which is "A structural SVM approach for reference parsing" [1]. This approach involves using structural support vector machines which is a supervised machine learning algorithm/method used for forecasting structured outputs. The feature extractions featured in "A structural SVM approach for reference parsing" involves preliminary processing and splitting the reference into individual word tokens based on criteria \cite{zhang2011structural}. They use these features to identify different entities that pattern match to their outputs. A second method proposes to avoid content based analysis due to their unreliability and instead proposes “a Web-based CME approach and a citation enriching system, called as BibAll, which is capable of correcting the parsing results of content-based CME methods and augmenting citation meta-data by leveraging relevant bibliographic data from digital repositories and cited-by publications on the Web” \cite{gao2012web}. CiteSeer, which is a citation indexing system that refers to a large freely view-able database of scholarly articles with citation and bibliography meta-data which updates automatically. This idea is referenced from a paper Cameron wrote on the topic \cite{cameron1997universal}. "Digital libraries and autonomous citation indexing" created a way in which authors didn't have to expend their energies to see their work in the database \cite{lawrence1999digital}. It instead grabs the papers from the web in pdf format, grabs the citation and its context then uploads it to a database in order to advance the ease of searching for literature and its assessment. Another proposed system in 2007 suggests an unsupervised system and “does not require or rely on patterns encoding specific deliminators of a particular citation style” \cite{cortez2007flux}. The model they propose involves 4 steps, splitting the string into its different element blocks.  “matching” an associated meta-data field with an element block. Takes unmatched element blocks and examines them for more associations based off their position within the citation. It collects all of the element blocks and chains them together when they are associated to the same meta-data field. Their results were as follows."the extraction quality obtained by FLUXCiM, even without user intervention, reached F-Measure levels above 92\%” \cite{cortez2007flux}.

\par
Template matching approaches uses syntactic pattern matching against a knowledge base of known templates. The citation is put against these templates to find the closest match, upon success the best fit template is used to label the tokenized version of the citation as fields\cite{councill2008parscit}. An authoritative example of this in action can be seen in ParaTools \cite{jewell2000paracite}. These perl modules hold an estimated 400 templates which are used to match against reference strings. These templates may encompass a large selection of references but it manifests scope problems where not all types of references are covered within these templates. This approach does not scale effectively due to the exercise of adding templates manually.

Yin \textit{et al} \cite{yin2004metadata} uses 'Bigram' HMMs to extract perform meta-data extraction. A HMM is a statistical Markov model, a finite state machine containing appropriate state transitions and symbol output/emissions. The process involves the production of a string of symbols by a starting state symbol, transitioning to other states, outputting symbols dictated by the current state of the model until it reaches a concluding state. Probability distributions are associated with each set of states over the symbols that are in the emission set and a probability distribution over the output transitions. A reference can be observed as a arrangement of fields such as 'Author', 'Title', etc. For a HMM, every state within can be marked as a label associated with one of the above fields mentioned. The extraction process proceeds by determining the arrangement of states that is most probabilistic for generating the complete reference and then placing corresponding labels with their fields according to the sequence of the states. Yin \textit{et al} \cite{yin2004metadata} uses a "dynamic programming solution called the viterbi algorithm" which solves the problem in\textit{
$O(TN^2)$} time. Another HMM by Hetzner \textit{et al} \cite{hetzner2008simple} describes a similar approach without the use of 'Bigrams'.
\par
\begin{figure}
\begin{center}
\caption{Simple Example of Hidden Markov Model}  \label{fig:HiddenMarkov}
\includegraphics[scale=0.50]{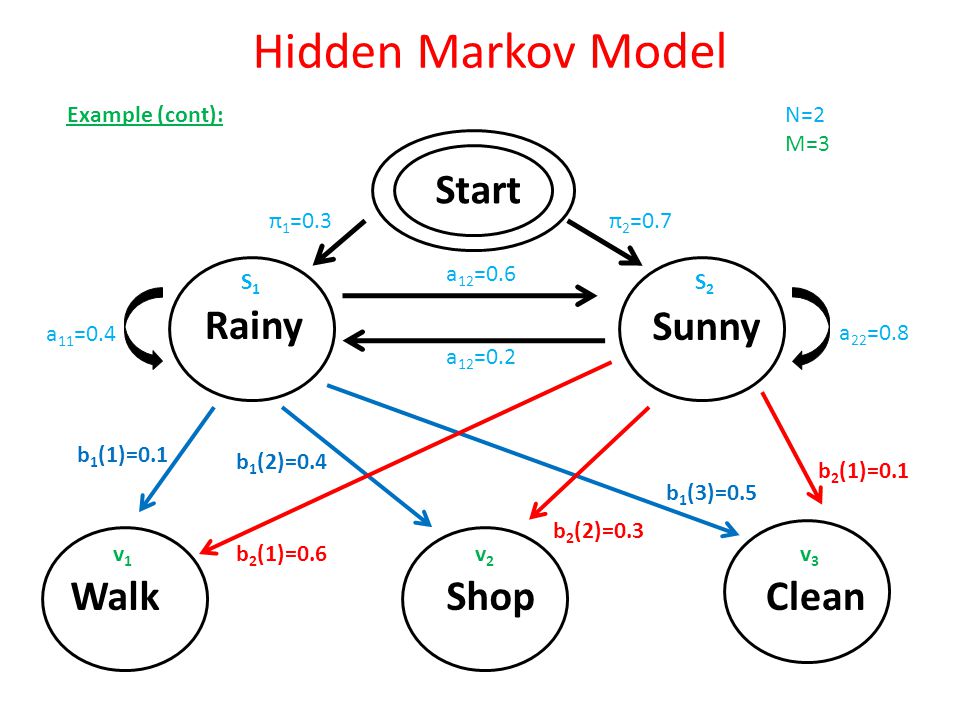}\\
Figure (\ref{fig:HiddenMarkov}) shows a simple Hidden Markov Model showing the different states with their attached probabilities for each transition as well as the outputs that are possible from each \cite{bradham}.
\end{center}
\end{figure}

 
A simple example can be seen in figure \ref{fig:HiddenMarkov}.

\par


\par
Within ParsCit \cite{councill2008parscit}, they use what is known as a condition random field (CRF) \cite{lafferty2001conditional} format to learn a model from an annotated reference data-set that can be utilized on hidden or otherwise unknown references.The ParsCit system works relatively well on a small evaluation data sets with good accuracy performance. This model of learning allows for adequate scaling in regards to 'conditionally dependent features' that may overlap. They use a variety of manually engineered features to improve on error prone past systems. Just to list a few of these features from ParsCit \cite{councill2008parscit}.
\clearpage
\begin{description}
\item Token identity (3): We encode separate features for each
    token in three different forms: 1) as-is, 2) lowercased,
    and 3) lowercased stripped of punctuation.
\item N-gram prefix/suffix (9): We encode 4 features for the
    first 1-4 characters of the token, similarly for the last
    1-4 characters. A single feature also examines the last
    character of the token, encoding whether it is uppercase,
    lowercase or numeric.
\item Orthographic case (1): We analyze the case of the token,
    assigning it one of four values: Initialcaps,
    MixedCaps, ALLCAPS, or others.
\item Punctuation (1): Similarly, we give fine-grained
    distinctions for the punctuation present in
    token: leadingQuotes, endingQuotes,
    multipleHyphens (occasionally found in page
    ranges), continuingPunctuation (e.g., commas,
    semicolons), stopPunctuation (e.g., periods, double
    quotes), pairedBraces, possibleVolume (e.g.,
    “3(4)”), or others.
\end{description}
\par

CRF is by and large the most effective and accurate method of meta data extraction. With GROBID \cite{DBLP:journals/ercim/LopezR15} having extremely strong results in precision and recall - "GROBID provided the best results over seven existing systems, with several metadata recognized with over 90\% precision and recall". In \cite{outofbox}, the top most accurate and performance heavy citation parsers were all in the category of CRF using small to medium sized data sets, in the range of 200,000 - 420,000 reference strings. "We leverage higher order semi-Markov conditional random fields to model long-distance label sequences, improving upon the performance of the linear-chain conditional random field model." \cite{DBLP:conf/jcdl/CuongCKL15}

\cite{DBLP:journals/ercim/LopezR15, DBLP:conf/jcdl/CuongCKL15, DBLP:journals/dlib/TkaczykTB15}

\subsection{Pre-Processing of documents}
Pre-Processing is important because of the crucial need for accuracy when calculating citation density, connections and making recommendations. It is needed because of errors obtained in a lot of databases because of human error upon citing and creating references and also through formatting issues which do not maintain the consistency of the documents different sections. Leaving citation parsers bewildered. It is needed for a large, public data set because of the current open source data sets which seem to be error prone in several manners, the need for accurate, precise data is growing, as all of the libraries and web of citation systems grows exponentially.
The normal setting for extracting these reference strings is through document formats such as PDF. PDFs can cause a lot of noise within these references in the PDF if the reference contains special characters or escape characters before being added. Finding these references is the first step in extracting them. "Given a plain UTF-8 text file, ParsCit finds the reference
strings using a set of heuristics. It begins by searching for
a labeled reference section in the text. Labels may include
such strings as “References”, “Bibliography”, “References
and Notes”, or common variations of those strings. Text is
iteratively split around strings that appear to be reference
section labels" \cite{councill2008parscit}. A lower bound is put on where a reference section label can appear to which it is considered to be accurate, this criterion is set to around 40\% by default. The last match of reference section labels is treated as the reference section to analyze. Reference strings are then extracted based on heuristics and citation markers such as square brackets or parenthesis.

\par

\par

\begin{figure}
\begin{center}
\caption{Shows Layout of a CRF System}  \label{fig:CRF}
\includegraphics[scale=0.45]{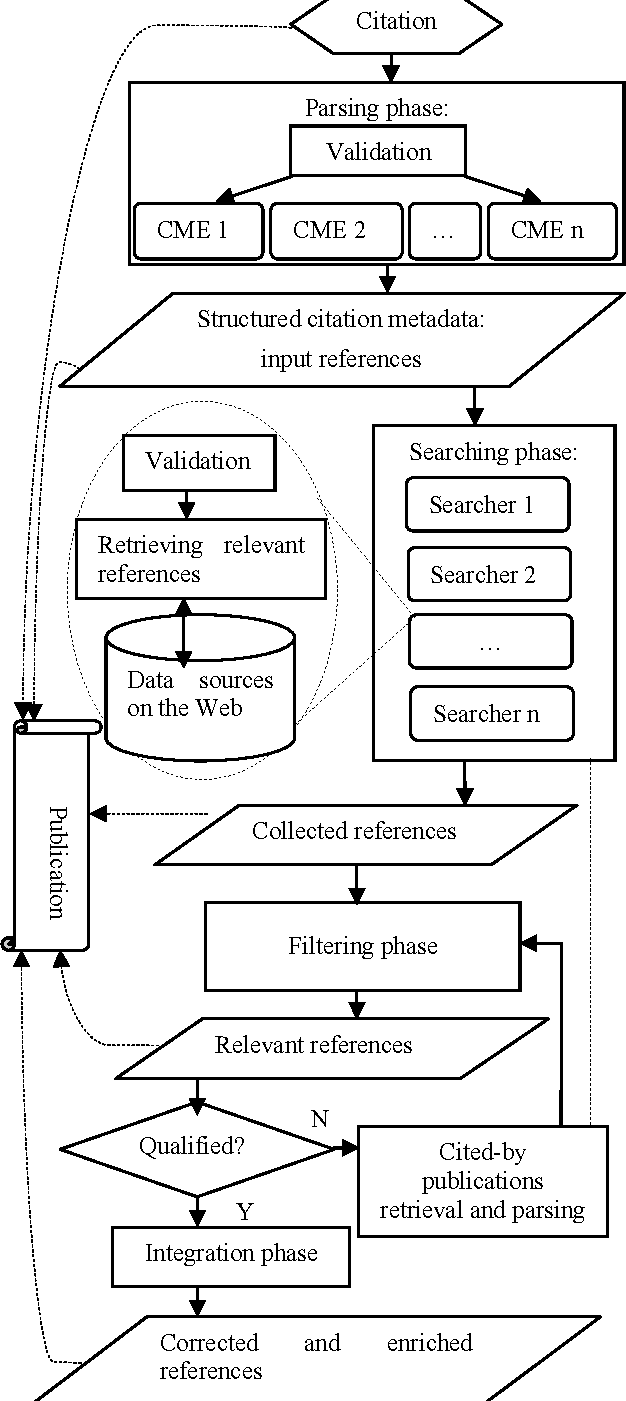}\\
Figure (\ref{fig:CRF}) Shows a complicated citation parsing system using conditional random fields \cite{gao2012web}.
\end{center}
\end{figure}


\clearpage
\par

In a supervised machine learning-based approach, reference
parsing is more formally expressed as a sequence tagging problem. In this type of problem, the input consists of a sequence of features, with the objective being to assign a corresponding chain of labels. This will bear in mind that it is not only the features that are a factor but also dependencies between the adjoining labels within the chain. Tokenization must be operated on the input reference string, this involves fragmentation. Each fragment is generally referred to as a token. Upon successful fragmentation, each token is appointed a label by the sequence tagger. Each label assigned usually coincides with desired meta-data fields and other special labels.Similar labels are concatenated to form the finished meta-data fields.

\par
Cermine tkaczyk \textit{et al} (2015) is another both supervised and unsupervised machine learning system for comprehensive meta data extraction. "The evaluation of the extraction
work flow carried out with the use of a large data set
showed good performance for most meta data types, with the
average F score of 77.5\%" \cite{tkaczyk2014cermine}. 

\par

Machine learning citation parsers perform better on average. This difference is largely due to recall.
"The average recall for ML-based tools (0.66) is three times as high as
non-ML-based tools (0.22). At the same time, the difference in
average precisions is small (0.77 for ML-based tools vs. 0.76 for
non-ML-based tools). The reason for this might be that it is
relatively easy to achieve good precision of manually developed
rules and regular expression, but it is difficult to have a high
enough number of rules, covering all possible reference styles."
\cite{outofbox}. Tkaczyk \textit{et al}

The reason a bigger data-set is in need, for training these machine learning based approaches instead of trying to find better algorithms or better machine learning models is because of the current trend in more data is always better. A famous quote from Peter Norvig a research director at Google asserts "We don’t have better algorithms. We just have more data.". Of course this is not always necessarily true, but in general more data is rarely a negative impact on results. A paper describing the means by which human complexities cannot be extrapolated through mathematical equations and instead accept those complexities, use what is useful and available. Data Halevy \textit{et al} (2009) \cite{halevy2009unreasonable}. The current data-sets available such as the cora data-set are of a reasonable size but nothing that could be considered massive \cite{seymore1999learning}. Especially in \cite{outofbox} they found pretty outstanding results which show major increases in significant field extraction accuracy's when using reasonably sized data-sets in the range of 100,000 unique references to 500,000. Due to these results, it indicates that by increasing the size of these data-sets to be massive. Performance of the models will also increase, be it a small amount of 2\% for well trained algorithms or much higher percentages in the high 20's for weaker ones. It will answer the question. Will a threshold be revealed on the performance of these models? Which indicates that better algorithms are the way forward in this area or that the upper bound may not exist on this problem and more data will just continue to make these citation parsers better.


\chapter{Methodology}
\section{Introduction to Scraping}
After trudging through the internet looking for data-sets that match my criteria and coming up empty. Automated scraping seemed like the most effective solution considering there were many sites that offer articles in BibTeX format with the click of a button. Scraping involves the requesting of html pages/files from servers, this usually involves making many requests to the same web page in order to grasp as much data as possible. This can also be referred to as DDOS if the frequency of requests gets high enough to interrupt the service of the website which happens to be illegal. A lot of systems are in place to stop possible DDOS'ing and we need to avoid those systems in order to allow our scraper to continue collecting for long periods. 99\% of the time, the system knows you are an AI and you are scraping but if you are "bashful" enough, it won't be worth the servers time to IP ban you as your are not worth their resources. To explain briefly what I mean by being "bashful" is referring to quite slow parameters for the scraper i.e. wait a rather long time between requests, use different media devices in headers per request, make it so it is not worth the sites time to ban you. There are a lot of pre-built scrapers for specific sites that can be used, alas mine are rather particular needs and I built my own scrapers in python. Scraping efficiently while not getting caught is a difficult balancing act and many elements affect the speed at which you collect.

\subsection{Security and Reliable Sources without inaccuracies}
All of the data scraped was freely available to the public and I don't need to consider data protection laws when porting it to a database. A main concept behind this investigation was making sure the data collected was of good standard from reliable sources so that the data did not contain errors or artificial entries. To achieve these goals of clean data, anytime I came across a suitable library I would manually check the accuracy of their \hologo{BibTeX} files through 10 trial runs. Which involved inspection of fields for errors in the text, such as syntax and noise. These were compared against other libraries which contain the same article to get a precise measure of their accuracy. These points are important to the overall goal, of supplying a large, open source data set because if the data is retaining inaccuracies it will not be useful when trying to train these parsing tools.
\par
\begin{figure}
\begin{center}
\caption{Shows number of lines scraped over 2-3 day interval data points} \label{fig:chartDataCollected}
\includegraphics[scale=0.40]{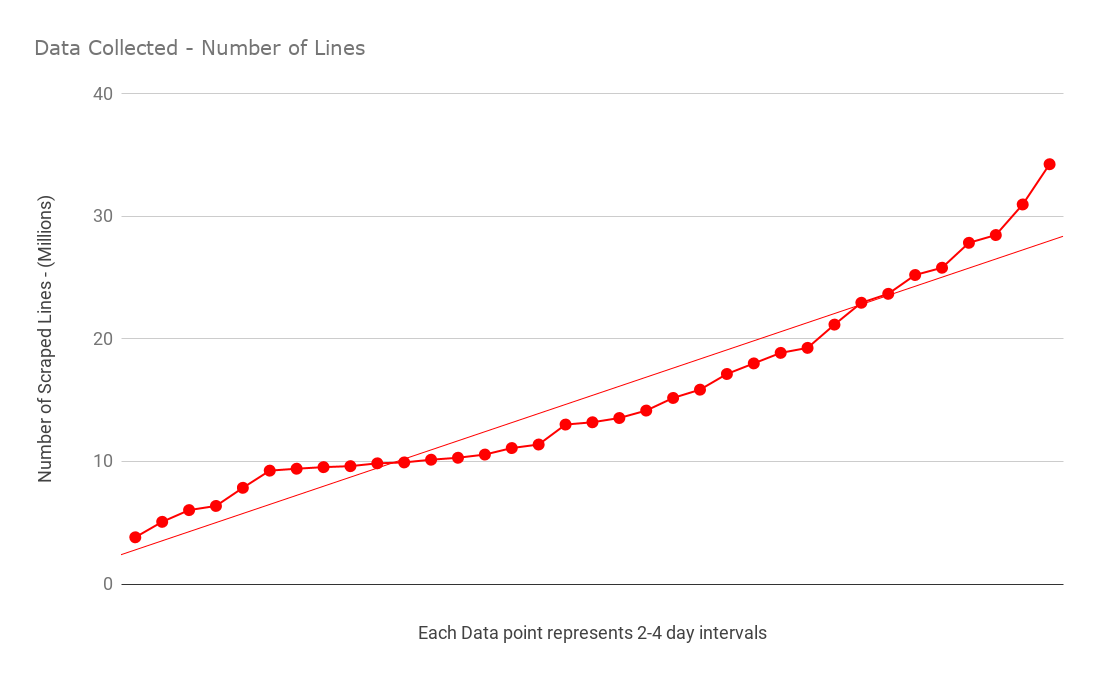}\\
Figure (\ref{fig:chartDataCollected}) shows 
\end{center}
\end{figure}
\par

\section{Finding efficiency} 
Google Scholar is a excellent resource for finding articles that suit your research tastes or articles that relate to other topics through citations. This is why scraping it seemed ideal for my data-set needs as a reliable product. Although upon further investigation citation counts were susceptible to manipulation by authors and spam alike \cite{Beel10c, Beel2010}. I underestimated the capabilities of Google's ability to track suspicious requests and block their requests. The structure in which they store their data on the site also meant that retrieving the \hologo{BibTeX} would take at least 3 requests for 1 return entry. This just made it an impossibility to scrape even when being 'bashful'. In a test scrape of Google Scholar I found that I was given the following image within 37 entries collected, which is overwhelmingly impressive. ACM was a honourable mention as it did indeed block my ip address on occasion.

\begin{figure}
\begin{center}
\caption{Google Captcha page upon requesting too fast} \label{fig:Google_Sorry_3}
\includegraphics[scale=0.5]{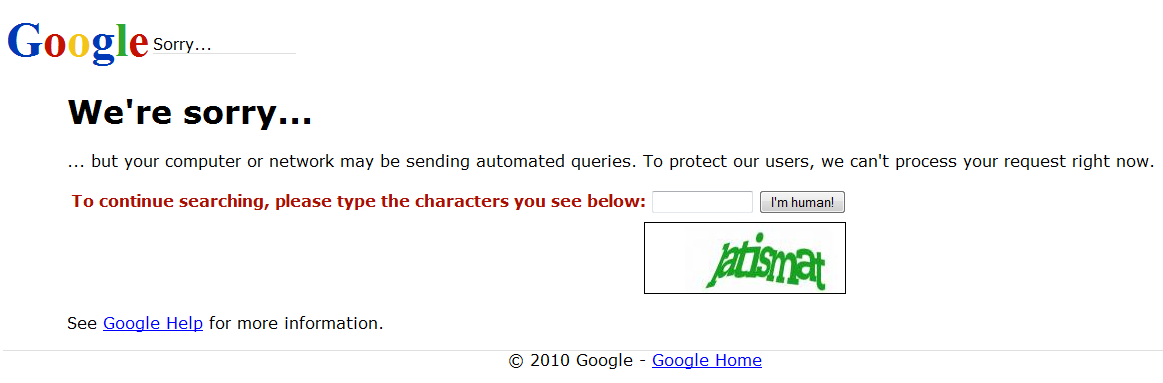}\\
\end{center}
\end{figure}

Captchas are a great way of checking for AI by giving the requester a non-automated readable question, mostly images that contain obscure looking characters, although I am sure there are machine learning models out there to solve these problems. An important aspect of the search for fast requests was on quite a basic but tedious level. How do I make sure I am getting at least one entry per request? To take the Google scholar example once more, The only way that I could iteratively and uniformly scrape entries was through 3 requests, 2 of which were for information which supplied the third. This was a ridiculous amount of overhead in contrast to the return I was receiving for this example. This lead to the search for efficiency through 1 request per one entry or better. This consecutively involved looking for hidden links/urls which were iterative and did not need more than 1 request. When I say iterative I mean does the link itself contain an identification number for each entry that can be iterated through with requests. If the ID's are random instead of uniform, this may cause the scraping of undesirable data.  \par

These are some examples of what I referred to as 'hidden links', only because it involved opening up developer tools in your favourite web browser and following the network sequence of requests and headers which needed to be analyzed to find the correct one.\\
'@' symbols indicate that a descriptive placeholder is in the example instead of a conformative example.
\clearpage
\begin{figure}
\begin{center}
\caption{Examples of link parameters and the iterative IDs associated}
\label{fig:ExamplesOfLinks}
\begin{lstlisting}
    ACM
            https://dl.acm.org/downformats.cfm?id=965498&parent_id=&expformat=bibtex
    
    CCS
        https://liinwww.ira.uka.de/csbib/ + query                   
            parameters['query','results','sort','start','maxnum']
    TeXMed
        http://www.bioinformatics.org/texmed/cgi-bin/list.cgi?PMID=29168941
    DBLP
            dblp.uni-trier.de/pers/tb2/@FirstNameBeforeComma:@NameAfterCommaSeparated
            ByUnderscores.bib
    MS
        https://academic.microsoft.com/cite/Bibtexs?pID=1577993811
    
    IEEE
        http://ieeexplore.ieee.org/xpl/downloadCitations?citations-format=citation-on
        ly&download-format=download-bibtex&recordIds=8110474
\end{lstlisting}
\end{center}
\end{figure}
\par





\subsection{DBLP}
On account of most of the scrapers having id's for each of their \hologo{BibTeX} entries, I was on average getting 1 \hologo{BibTeX} entry per 1 request which is good depending on the speed of your requests and if you can cut the time down by running things in parallel or concurrently. DBLP, which is a computer science bibliography library, had the same indexing system, but I needed results faster than the '1 for 1' which was occurring for the other scrapers mainly due to time constraints for the project. After some digging, I found that if you got an authors details you could download their articles as \hologo{BibTeX} or scrape it from a view they provided. This increased the productivity and efficiency of the scrapers by quite a large range considering it was now 1 - N returns per request. 
\par
\begin{figure}
\begin{center}
\caption{Shows progression of lines gathered for each individual scraper} \label{fig:chartProgression}
\includegraphics[scale=0.55]{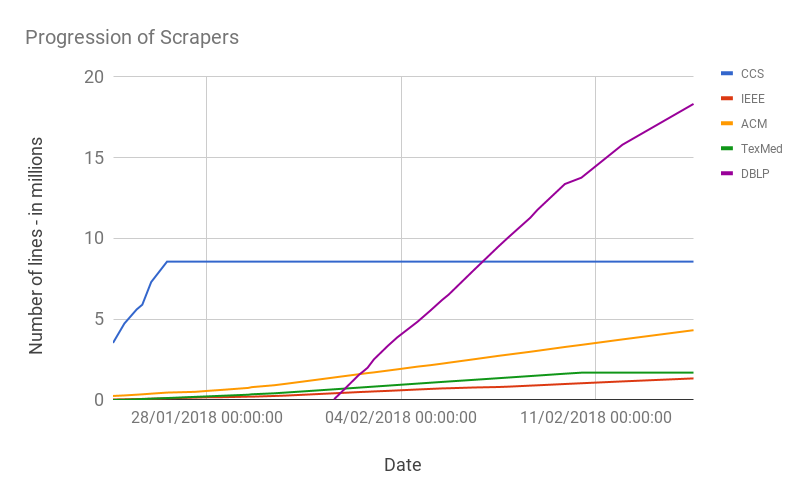}\\
Figure (\ref{fig:chartProgression})  
\end{center}
\end{figure}
\begin{figure}
\begin{center}
\caption{Shows progression of lines gathered using scrapers. It is also normalized by the most efficient data point.} \label{fig:chartOfEfficiency}
 \includegraphics[scale=0.4]{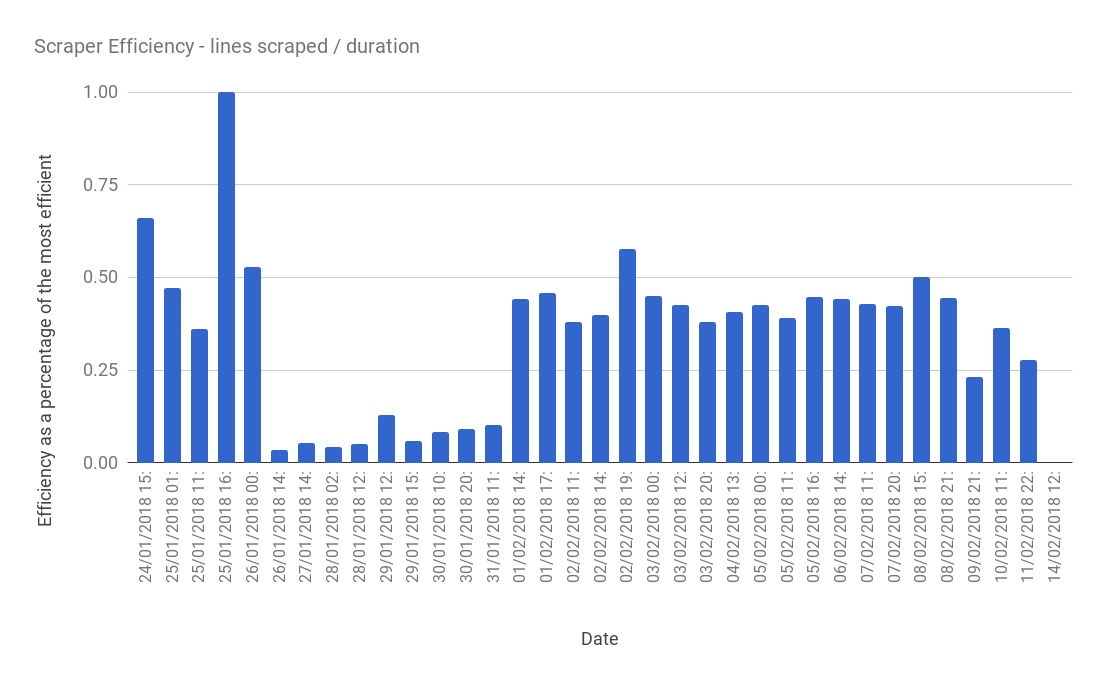}\\
Figure (\ref{fig:chartOfEfficiency})  
\end{center}
\end{figure}

In figure \ref{fig:chartProgression} and figure \ref{fig:chartOfEfficiency}, it is clear that they correlate and explain each other rather well. The high starting efficiency's in \ref{fig:chartOfEfficiency} are because of the output of CCS which had rather effective requesting and a relatively low request delay. This eventually drops off because the scraper finished incredibly fast and had nothing else to collect. Then as soon as DBLP begins , the efficiency's go up to a steady amount until the end of the entire data gathering section.

\clearpage
\par


\par

\clearpage

\section{Scraping under the radar}
Automated scraping using your own 'bot' violates almost every single search engine's terms of service especially if your bot doesn't even look at their robot.txt file which outlines the terms and rules that bots must follow when using the site.
"The search engines aren't naive. Google knows every search engine operator scrapes its results. So does Bing. Both engines have to decide when to block a scraper"\cite{searchnewcentral_2017}. After some testing the equation of decisions of whether to block or not block a scraper becomes clear. 
\begin{itemize}
  \item The potential load put on the server by the scraper
  \item The potential load put on the server because it needs to block the scraper
\end{itemize}
\cite{searchnewcentral_2017}
It basically comes down to being on the right side of these two parameters. Even if you are inevitably found out as a bot, your scraper should be so unobtrusive that it is not worth the sites time to waste CPU cycles on you.
Humans don't click on the different links within milliseconds of each request at a speed that is almost unfeasible. Making the scrapers as human as possible increases the probability of not getting stopped or IP banned.The main two tune-able parameter that each of my scrapers used to disguise themselves are. TD - 'Threshold Delay' and RID - 'Random Incremental Delay'. TD is a delay that is constant and stops the scraper from making requests too quickly in succession. RID is to give the scraper random elements of time delay to give the illusion of being a human. These parameters are tune-able if you have a server which can dynamically change its IP address, a tunneled VPN or some way of avoiding a ban if you get caught too many times. The tuning process involves taking in scraper information as well as looking at logs of the interactions of your scrapers and altering the appropriate parameter within the scraper. furthermore, using a random user agent within the header files of each request from a large array of devices can increase your probability of going undetected drastically.



\section{Maintenance}
On my server I had upwards of five to seven scrapers running at any one time. This meant that I was constantly checking their progress and fixing bugs that would arise quite often. I implemented a simple tracking system so that if any of the scrapers was cut short on execution, it would give information on the error, the id of the last URL that was accessed and the current amount of entries scraped. This allowed for easily accessing, debugging and resetting of the scraper to pick up where it left off. With a small server it can be difficult to organize resources to be as efficient as possible, 

\section{Parsing \& Removal of Impurities}
\hologo{BibTeX} is a format which is relatively simple to parse in most languages. The issue lies in entries containing special characters which will affect a parsers interpretation of an entry, which in turn will produce errors. There exists many 'Bib cleaners' which attempt to clean \hologo{BibTeX} databases by removing duplicates, checking that the entries contain the correct fields for that type of entry and call attention to infractions in the bibliography's which may affect parsing. Random manual inspection can be useful for getting a 'taste' for what the data consists of and if there may be possible problems with a conversion. A field within the DBLP scraper data contained syntax that was reserved for \hologo{BibTeX} entries which would affect the conversion when a parser made attempts on it. Since the field was not one of major concern such as 'author' or 'title' I removed the field from ever entry it existed in to alleviate the issue.    

\subsection{The 'HomePage' problem}
Within the DBLP scraper I noticed a large number of similar disparities which had accumulated due to the @misc field type for \hologo{BibTeX} entries.All \hologo{BibTeX} field types can be seen in the B appendix. Misc in particular is the shortening of miscellaneous and is used when the type of the entry is not apparent. Within DBLP there contains thousands of entries for people who have homepages in \hologo{BibTeX} format which does not contain useful fields for the training these algorithms, especially considering the misc type does not require any fields to be valid. 



\chapter{Conversion of BibTeX \& Building the Data-Set}

\section{Citation Styles}

When referencing research papers and other scholarly articles, particular styles are used depending on the preference of the university or department. This is why citation parsing for meta-data generation is such a complex issue, of course if the parser knew which citation style is being used it could reverse engineer the reference to get back the correct fields. This is because of the way in which references are created, the details of each style are contained in CSL files. Which are publicly available at \cite{CSL}. These CSL files contain an XML schema for each style that can be used to convert \hologo{BibTeX} to a bibliographic reference which can be inserted into the paper. For each \hologo{BibTeX} entry that was scraped, that information is used alongside each of the thousand or so CSL styling files in order to generate appropriate bibliographic reference strings. Meaning that a single \hologo{BibTeX} entry within the database will contain citation information for every style that was used in the conversion. These bibliographies are needed so that base truth values can be obtained when evaluating the 'out-of-the-box' results, these results will be put in contrast to the trained versions of the tools.


\section{Annotating Citation References}
Annotated Citation References are references which contain meta-data fields in XML format which outline the fields that belong to the associated reference tokens. In order to create these annotated references with the correct corresponding fields or labels, the CSL files need to be altered in order to make the additions of the XML formatted tags which encompass the different tokens of the reference string. Manually altering over a thousand CSL files to reflect these changes is unfeasible, and such is why I begin writing scripts to automate this problem. One of the main problems with automating this process is the need for the tags to be accurate and since these CSL files are not always uniform in their layout XML, it makes for quite a difficult scheme of finding where to place tags within it. In order to write scripts for this, I chose to copy the appropriate tag name of what the CSL document was referring to that segment as, and add it to the correct position within the XML tags. This is why the uniformity of the field labels of the \hologo{BibTeX} analogous to the labels of the annotated references was such an issue.  
\par 
\bigskip
\begin{lstlisting}
<author><surname>Shachter</surname> <firstname>R. D.</firstname>, <surname>Levitt</surname> <firstname>T. S.</firstname>, <surname>Kanal</surname> <firstname>L. N.</firstname> & <surname>Lemmer</surname> <firstname>J. F.</firstname> (eds)</author> <issued> 1990. </issued> <edition>UAI ’88: Proceedings of the Fourth Annual Conference on Uncertainty in Artificial Intelligence, Minneapolis, MN, USA, July 10-12, 1988</edition><publisher>North-Holland</publisher> .
\end{lstlisting}

\section{CiteProc Software \& Efficiency Challenges}
CiteProc-java is a Citation Style Language (CSL) processor for Java, it interprets and translates CSL styles and generates citations and bibliographies \cite{michel}. Citeproc uses a CSL processor in order to make the conversion to a citation. Initializing the processor takes the style as a parameter, which means that you cannot use the same processor and change the style parameter for each case. This brings about massive inefficiency challenges as well as questions relating to the order of operations to be carried out for the conversion. There are two ways of looping the order of execution, one option is to initialize every processor for every style that is needed for conversion and iterate through each \hologo{BibTeX} entry object. For each entry, use the processor of each style to convert to its associated citation/reference. This will include the altered Annotated CSL files as well. This way, although it is expensive to initialize all of the processors in the beginning and each of the processors takes up a large chunk of RAM each, since the same object is being updated with the citations values it means that it will most likely be cached and it will be faster than having to continually be pulling a new object from the database for each style insertion. The other approach is for each style, iterate through each object and make the conversion. With this approach it means that initializing each style processor may be expensive, it means that the overall amount of RAM used at any given time is reduced by the amount of style processors. After testing both approaches with one iteration, I found that my server did not have enough RAM to hold all of the CSL processors at once. This forced the first approach to go into action, as the latter was unfeasible in the current state of affairs of my server. 

\par

The \hologo{BibTeX} text files needed to be split into smaller subsets of files because when the conversion process begins, it reads the \hologo{BibTeX} file and parses the entire file putting it into a \hologo{BibTeX} database object in RAM. If the file was too large, memory leaks would occur which is why they needed to be split. This became an advantage anyway, due to the fact it made it easier to write parallel code for completing the conversion.  

\section{Structure of the Data-set}
The structure of the data-set is dictated by the fields and information needed in order to train and evaluate existing citation tools which are lacking in regards to the extensiveness and diversity of its training data. The \hologo{BibTeX} information at the top of each entry is what is used to create the reference strings which are contained in the 'Citation' array, each of a different style. The 'bibRef' is short for bibliographic reference and is the original unaltered reference. The 'AnnoRef' is the annotated version of the original reference which will be used for training a model with the currently available machine learning tools. The Data-set itself is in the form of csv/json, the data was first stored in a Mongo database.
\clearpage
This shows the schema of the database, with "..." indicating more values of similarity continuing and quotes indicating values of fields. \\
\par 
\begin{lstlisting}
// One Entry in JSON format
{ 
    // BibTeX Information 
    _id     :   "   ",
    Title   :   "   ",
    Author  :   "   ",
    Volume  :   "   ",
    Pages   :   "   ",
    ...
    // Citation/Reference Information
    Citation : [    
        {
            Style       :   "   ",
            bibRef      :   "   ",
            AnnoRef     :   "   " 
        }
        {
            ...
        }
        ...
    ]
}
\end{lstlisting}

\section{Meta-Data Field consistency \& Noise}
The need for clean, consistent and clear meta-data fields are necessary when making string comparisons between both the meta-data fields themselves and the inner values of the fields. This means that inaccuracies of meta-data fields can impact assessment percentages of parsing tools. A simple example of this is if the \hologo{BibTeX} entry contains a field "Pages" and when the entry is converted to a annotated citation the entry for "Pages" has gotten a different field name from the original such as "Locality".

\section{Results}
The overall scraping resulted in 7-8 text files containing an estimated 2.5 million \hologo{BibTeX} entries in total. Which when multiplied by the number of styles I have prepared for annotation (1600) gives 4 billion possible instances over 2.5 million unique reference strings
Currently the data-set contains approximately 20 million instances of 400,000 unique reference strings which were converted using unique styles. Only 50 styles were used in the conversion because of the way in which the data was being converted, which I mentioned in this chapter. Since one style had to go through every \hologo{BibTeX} entry in that particular file, It meant that currently only 50 styles have been used in total for the instances of references in the data set. Putting the order of execution in the opposite fashion was unfeasible, but would of used all 1600 styles on one entry at a time.
7-8GB of RAM was being chunked away at in order to have concurrent updating on different ports. Between the Data that was stored from \hologo{BibTeX} entries, code, resources for styles/DBconnectors and the data set entries, Over 70GB of disk was being used up. Until the point of when errors appear because tab completion could not execute due to lack of memory.
A confusion matrix is regularly used to depict the performance of a machine learning model. Precision is the ratio of correctly predicted positive observations to the total predicted positive observations, high precision rates correlates with a low false positive rate. The question that precision answers is, of the fields that we labeled, how many of those that were labelled were correct? Recall, also referred to as sensitivity is the ratio of correctly predicted positive observations to all the observations within the classification class. recall explains the question, of the ground truth fields, how many did the tools output were labeled. 
F1 score is the weighted average of both precision and recall. This can be better than accuracy if the false positives and false negatives rates are drastically different to each other. Accuracy works best when false negatives and false positives are relatively close in score or have a similar cost. This is also referred to as having a 'even class distribution'.
\par
In table \ref{table:FieldsBibTeX}, This shows the number of fields overall contained in each of the scrapers \hologo{BibTeX} entries. CCS and DBLP appear to have the most diverse fields, with almost no 0's contained in each. IEEE and TeXMed appear to have very focused sets of data, which when visually inspected is absolutely true, with the majority of their \hologo{BibTeX} entries being exactly 10 lines containing the exact same fields for each. The sheer size difference of CCS,DBLP and the rest of the scrapers may be a more in tune reason for the diversity but manual inspection reveals the consistency of IEEE and TexMed \hologo{BibTeX} 

\par 
In table \ref{table:TypesBibTeX}, This shows the types of \hologo{BibTeX} entries that were gathered from each scraper.

\clearpage
\begin{table}[h!]
\begin{center}
\caption{Number of Fields in \hologo{BibTeX} Entries for each Scraper}\label{table:FieldsBibTeX}
\begin{tabular}{||c|c|c|c|c|c||}
\hline
& ACM & CCS & IEEE & texmed & DBLP \\ 
\hline
address & 233790 & 21897 & 0 & 0 & 51 \\ 
\hline
annote & 0 & 21681 & 0 & 0 & 0 \\ 
\hline
author & 323633 & 390186 & 143981 & 173744 & 701802 \\ 
\hline
booktitle & 63273 & 161895 & 105049 & 0 & 401825 \\ 
\hline
chapter & 24556 & 186 & 0 & 0 & 17 \\ 
\hline
crossref & 0 & 45388 & 0 & 0 & 0 \\ 
\hline
edition & 12 & 537 & 0 & 0 & 15 \\ 
\hline
editor & 51950 & 66112 & 0 & 0 & 199695 \\ 
\hline
howpublished & 0 & 408 & 0 & 0 & 0 \\ 
\hline
institution & 0 & 5355 & 0 & 0 & 60 \\ 
\hline
journal & 233794 & 186669 & 65558 & 173744 & 294255 \\ 
\hline
key & 1452 & 2286 & 0 & 0 & 335 \\ 
\hline
month & 210522 & 78808 & 168872 & 128940 & 20 \\ 
\hline
note & 13447 & 36114 & 0 & 0 & 47 \\ 
\hline
number & 232433 & 155326 & 64793 & 164979 & 227003 \\ 
\hline
organization & 0 & 1557 & 0 & 0 & 175 \\ 
\hline
pages & 295327 & 281575 & 167902 & 173744 & 636320 \\ 
\hline
publisher & 353843 & 137330 & 0 & 0 & 341704 \\ 
\hline
school & 0 & 14838 & 0 & 0 & 4575 \\ 
\hline
series & 50239 & 30282 & 0 & 0 & 94151 \\ 
\hline
title & 393858 & 557415 & 275656 & 173744 & 1050427 \\ 
\hline
type & 1 & 17166 & 0 & 0 & 200 \\ 
\hline
volume & 232975 & 210201 & 82582 & 165769 & 387578 \\ 
\hline
year & 353860 & 388518 & 170607 & 173744 & 1050460 \\ 
\hline
\end{tabular} 
\end{center}
\end{table}
\par

\begin{table}[h!]
\begin{center}
\caption{Types of \hologo{BibTeX} Entries for each Scraper}\label{table:TypesBibTeX}
\scalebox{0.9}{
\begin{tabular}{||c|c|c|c|c|c||}
\hline
& ACM & CCS & IEEE & TexMed & DBLP \\ 
\hline
article & 233790 & 200694 & 65664 & 173123 & 294244 \\ 
\hline
book & 16176 & 9762 & 0 & 0 & 5310 \\ 
\hline
booklet & 0 & 21 & 0 & 0 & 0 \\ 
\hline
conference & 0 & 63 & 0 & 0 & 0 \\ 
\hline
inbook & 0 & 126 & 0 & 0 & 0 \\ 
\hline
incollection & 24556 & 5826 & 0 & 0 & 5130 \\ 
\hline
inproceedings & 63273 & 113877 & 105058 & 0 & 396695 \\ 
\hline
manual & 0 & 528 & 0 & 0 & 0 \\ 
\hline
mastersthesis & 190 & 6690 & 0 & 0 & 0 \\ 
\hline
misc & 0 & 36024 & 0 & 0 & 0 \\ 
\hline
phdthesis & 12473 & 5256 & 0 & 0 & 4371 \\ 
\hline
proceedings & 2208 & 3288 & 0 & 0 & 344676 \\ 
\hline
techreport & 1185 & 6798 & 0 & 0 & 0 \\
\hline
unpublished & 0 & 708 & 0 & 0 & 0 \\ 
\hline
\end{tabular}
}
\end{center}
\end{table}

\clearpage
In figure \ref{fig:FieldsScraper}, it displays the number of each field within each scraper. The most fields gathered are the most important for citation impact calculations as well as recommendations for other scholarly articles usually, although the fact that "year" is higher than author, leads me to believe their may be a lot of conferences within DBLP which do not contain authors but rather event details. Everything else seems to be as expected with "pages" , "journal", "volume" "publisher" and "number" all having medium to high results.
\par
In figure \ref{fig:TypesScraper}, it depicts the number of types of \hologo{BibTeX} entries within each scraper. The majority of the dataset is made up of "articles", "inproceedings" and "proceedings". I believe the large amount of "proceedings" in DBLP relates to the lack of "author" fields that were collected in comparison to the "year" field.
\par
\clearpage

\begin{figure}
\begin{center}
\caption{Number of Fields in \hologo{BibTeX} entries, distributed over the different scrapers}  \label{fig:FieldsScraper}
\includegraphics[scale=0.35]{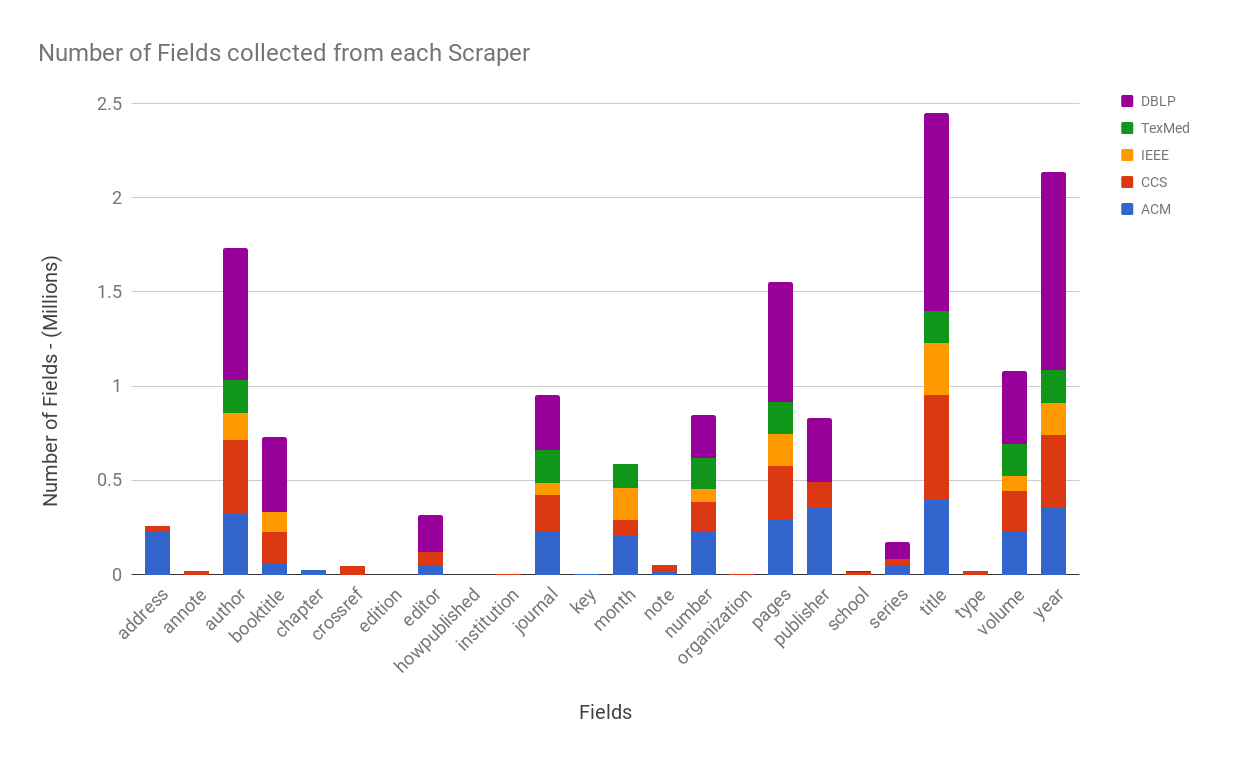}\\
Figure (\ref{fig:FieldsScraper})  
\end{center}
\end{figure}

\begin{figure}
\begin{center}
\caption{Number of Types of \hologo{BibTeX} entries, distributed over the scrapers}  \label{fig:TypesScraper}
\includegraphics[scale=0.350]{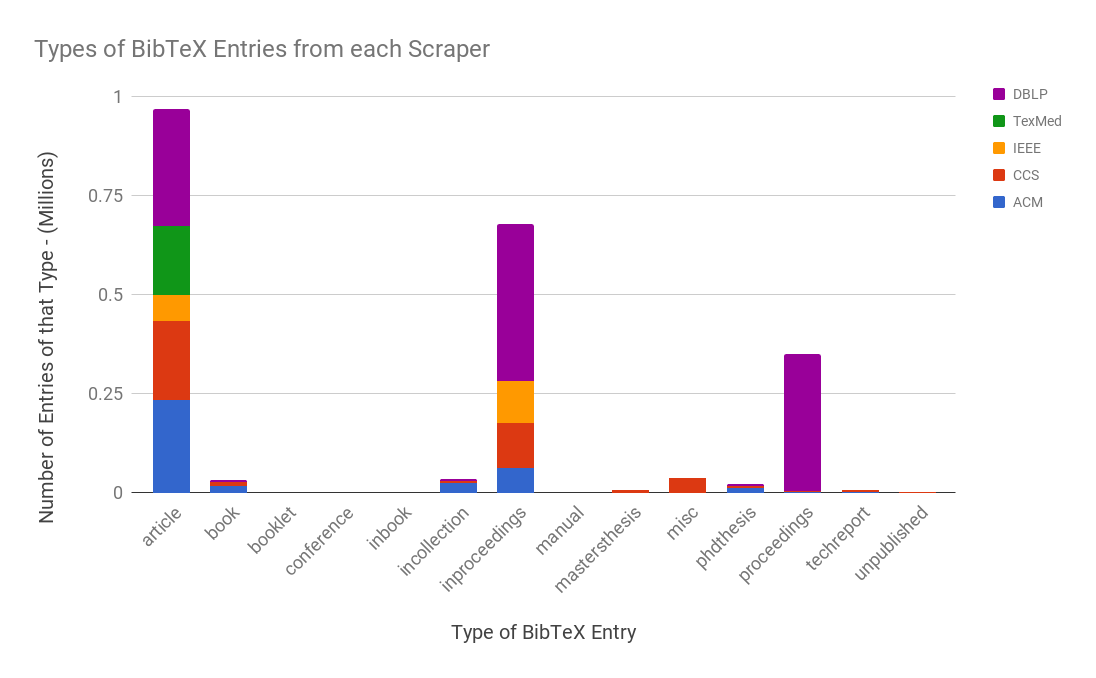}\\
Figure (\ref{fig:TypesScraper}) 
\end{center}
\end{figure}

\clearpage


    \chapter{Future Work \& Data-Set Usability}
This Data-set will be used for training citation parsing tools which feature machine learning based approaches. The tools will learn custom parsing rules from this training data. To mention a few of the open sources tools, CERMINE \cite{DBLP:journals/ijdar/TkaczykSFDB15} , GROBID \cite{lopez2009grobid}, ParsCit \cite{councill2008parscit}. 
\section{The Process of Evaluation Testing}
'Out of the box' versions of the tools contain pre-trained machine learning models which are used if no other model is provided. The comprehensive data-set is split into 66\% training data and 33\% evaluation sets. A lot of the tools ask for the annotated references to be formatted in a particular way before being processed. Usually it is just newline delimited. First the evaluation section of the set must be tested using each tool to get initial values for 'out of the box' performance. These results will be put in comparison to later results from the trained models using confusion matrices or basic accuracy percentages. The same tool is then trained using the training data and their accuracy's are put into contrast in regards to "precision" , "recall" and "F1 score".
\par


\par
\clearpage
\begin{figure}
\begin{center}
\caption{ Shows the dissection and pattern matching of a typical research article }  \label{fig:FieldsScraper}
\includegraphics[scale=0.7]{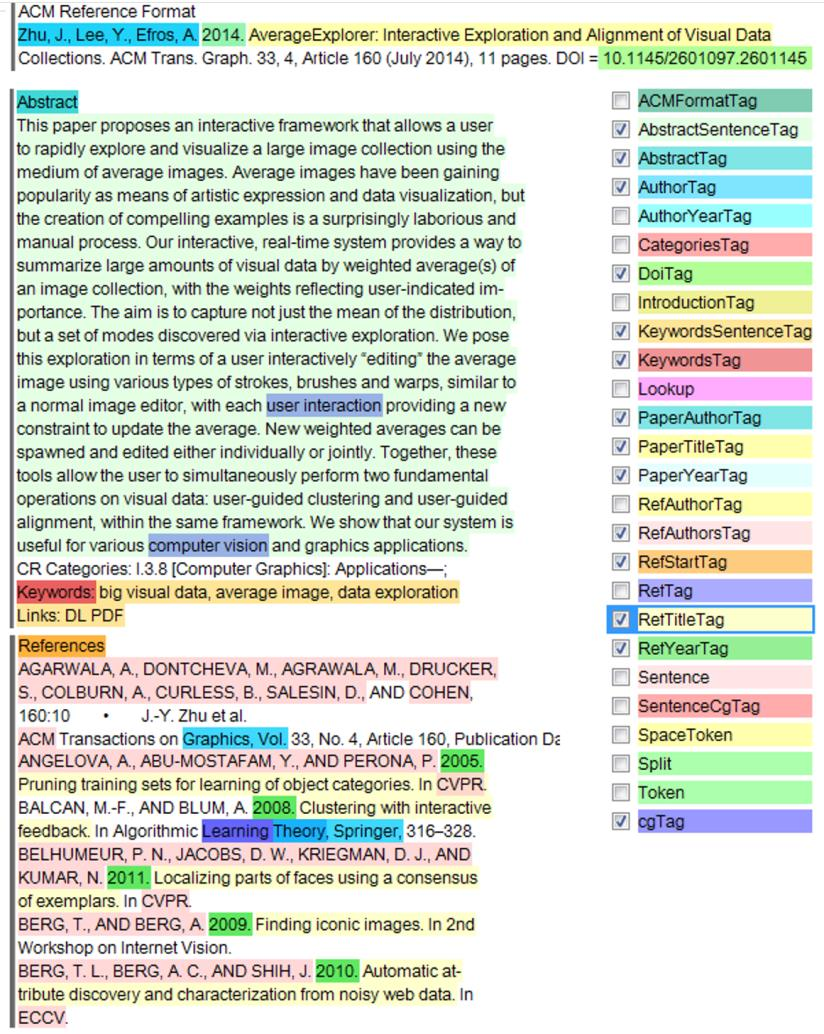}\\
Figure \ref{fig:FieldsScraper} depicts a visual representation of the tagging and pattern matching that happens when parsing a scientific research document, showing the field labels on the right legend \cite{unknown}.  
\end{center}
\end{figure}
\clearpage
\par
\subsection{Comparison Method}
For each tool, the given output fields that were extracted for each reference are compared against the annotated values of the same reference stored within the data-set which are referred to as the 'ground truth' values. The output fields are altered slightly in order to clean them, which includes cleaning of special characters and escape characters, conversion to lowercase and adjusting of some intermittent characters such as hyphens. After subsequent cleaning, true and false values are attached to each extracted field based on string comparisons between cleaned extraction fields and ground truth values. In order for an extracted field to be considered correct its type and value must be equal to one of the fields in the ground truth references. Removal of edge punctuation and variable elements of the string such as commas and ampersands can help ease the comparison process, we don't particularly care if the style uses and \& or and as that is not what is important within the extraction process.
\newline
\par 

\includegraphics[scale=0.6]{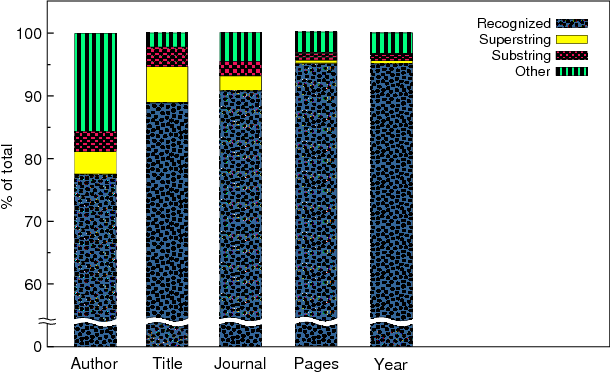}
\par
figure (6) - \cite{tkaczyk2014cermine} - Shows string comparison results relating to the length and manner of each as a percentage.

In figure (6), recognized means that the strings were equal and a 1 was assigned. Superstring implies that the tools output value contained the ground truth value but also contained extra elements not contained in the ground truth value. Substring implies the tools output value was contained in the ground truth value, but it was not the complete or entirety of the ground truth string.


\par
Another interesting way of finding the accuracy's of the outputted meta-data from the tools can be used by implementing a Levenshtein distance approach alongside a reasonable threshold being applied to get a correct or incorrect output. The reason for using the word interesting being that upon inspection of accuracy results, it can indicate or otherwise pinpoint issues with the tools inaccuracies which may relate to particular styles. It can also illustrate the sections of the reference which is being mistake based of the percentages given for the accuracy of each field extracted.

\section{Outlook}
The scraping was cut short because of time constraints on the project. Additionally the scarcity of available disk and memory on my server was a contributing factor.
My server was running out of disk due to the scale of these entries and the amount of data that was being stored in text files from the \hologo{BibTeX} entries. I underestimated the amount of RAM that Mongo would use in order to make fast updates to the DB and thus I could not concurrently run 4 of the conversions at the same time and eventually couldn't even run 1 DB without having memory issues. Since their are over 2 million \hologo{BibTeX} entries, at least 1600 configured styles for conversion, this brings the number of instances of converted reference strings up to 4 billion. In future, I would need to gather enough concurrent resources to make these conversions faster and complete. Especially under the time constraints it was rather difficult to make the conversions go any faster considering it was converting, updating and storing 10 entries per second. Which even under the hindrance of having a small server and concurrently converting with 4 instances, still comes out to about 1000 days to convert the entire \hologo{BibTeX} set.



\section{Conclusion}
Through research of countless scientific papers in this area and maneuvering through the struggles of scraping, I can see why open source data sets for citation parsing tools are sparse. Regardless of this, I am positive that this data set will continue to grow and increase performance for open source meta data extraction tools based on results from \cite{outofbox}, showing that retraining already trained tools can give promising results even with moderately sized data sets. Since this data set will become much larger over time, I predict that performance will grow with its size, until a threshold of reasonable uncertainty and the need for artificial noise and human error will be needed in order to train these tools further.






\addtocontents{toc}{\vspace{2em}} 

\appendix 

\chapter{Terms} 
%
%

\setstretch{1.5}  

\hologo{BibTeX}     -    Bibliography Type Setting\\
DBLP        - The dblp computer science bibliography is the on-line reference for open bibliographic information on computer science journals and proceedings\\
CCS     - The Collection of Computer Science Bibliographies\\
PubMed      - PubMed is a free search engine accessing primarily the MEDLINE database of references\\
TexMed      - a BibTeX interface for PubMed\\
ACM     - The Association for Computing Machinery is an international learned society for computing\\
IEEE        - The Institute of Electrical and Electronics Engineers is a professional association


%

\chapter{Information on BibTeX}

\textbf{ These are the standard entry types of BibTeX entries and the details of what is required or optional for each }
\begin{itemize}
\item article\\
An article from a journal or magazine. Required fields: author, title, journal, year. Optional fields: volume, number, pages, month, note.
\item book \\
A book with an explicit publisher. Required fields: author or editor, title, publisher, year. Optional fields: volume or number, series, address, edition, month, note.
\item booklet \\
A work that is printed and bound, but without a named publisher or sponsoring institution. Required field: title. Optional fields: author, howpublished, address, month, year, note.
\item conference\\
The same as INPROCEEDINGS, included for Scribe compatibility.
inbook
A part of a book, which may be a chapter (or section or whatever) and/or a range of pages. Required fields: author or editor, title, chapter and/or pages, publisher, year. Optional fields: volume or number, series, type, address, edition, month, note.
\item incollection\\
A part of a book having its own title. Required fields: author, title, booktitle, publisher, year. Optional fields: editor, volume or number, series, type, chapter, pages, address, edition, month, note.
\item inproceedings\\
An article in a conference proceedings. Required fields: author, title, booktitle, year. Optional fields: editor, volume or number, series, pages, address, month, organization, publisher, note.
\item manual\\
Technical documentation. Required field: title. Optional fields: author, organization, address, edition, month, year, note.
\item mastersthesis\\
A Master's thesis. Required fields: author, title, school, year. Optional fields: type, address, month, note.
\item misc\\
Use this type when nothing else fits. Required fields: none. Optional fields: author, title, howpublished, month, year, note.
\item phdthesis\\
A PhD thesis. Required fields: author, title, school, year. Optional fields: type, address, month, note.
\item proceedings\\
The proceedings of a conference. Required fields: title, year. Optional fields: editor, volume or number, series, address, month, organization, publisher, note.
\item techreport\\
A report published by a school or other institution, usually numbered within a series. Required fields: author, title, institution, year. Optional fields: type, number, address, month, note.
\item unpublished\\
A document having an author and title, but not formally published. Required fields: author, title, note. Optional fields: month, year.
\end{itemize}
\cite{bibtexHelp}

\textbf{}

\begin{itemize}
\item address\\
Usually the address of the publisher or other type of institution. For major publishing houses, van Leunen recommends omitting the information entirely. For small publishers, on the other hand, you can help the reader by giving the complete address.
\item annote\\
An annotation. It is not used by the standard bibliography styles, but may be used by others that produce an annotated bibliography.
\item author\\
The name(s) of the author(s), in the format described in the LATEX book.
\item booktitle\\
Title of a book, part of which is being cited. See the LATEX book for how to type titles. For book entries, use the title field instead.
\item chapter \\
A chapter (or section or whatever) number.
\item crossref\\
The database key of the entry being cross referenced.
\item edition\\
The edition of a book--for example, ``Second''. This should be an ordinal, and should have the first letter capitalized, as shown here; the standard styles convert to lower case when necessary.
\item editor\\
Name(s) of editor(s), typed as indicated in the LATEX book. If there is also an author field, then the editor field gives the editor of the book or collection in which the reference appears.
\item howpublished\\
How something strange has been published. The first word should be capitalized.
\item institution\\
The sponsoring institution of a technical report.
\item journal\\
A journal name. Abbreviations are provided for many journals; see the Local Guide.
\item key \\
Used for alphabetizing, cross referencing, and creating a label when the ``author'' information (described in Section 4) is missing. This field should not be confused with the key that appears in the cite command and at the beginning of the database entry.
\item month\\
The month in which the work was published or, for an unpublished work, in which it was written. You should use the standard three-letter abbreviation, as described in Appendix B.1.3 of the LATEX book.
\item note \\
Any additional information that can help the reader. The first word should be capitalized.
\item number\\
The number of a journal, magazine, technical report, or of a work in a series. An issue of a journal or magazine is usually identified by its volume and number; the organization that issues a technical report usually gives it a number; and sometimes books are given numbers in a named series.
\item organization\\
The organization that sponsors a conference or that publishes a manual.
\item pages\\
One or more page numbers or range of numbers, such as 42-111 or 7,41,73-97 or 43+ (the `+' in this last example indicates pages following that don't form a simple range). To make it easier to maintain Scribe-compatible databases, the standard styles convert a single dash (as in 7-33) to the double dash used in TEX to denote number ranges (as in 7-33).
\item publisher\\
The publisher's name.
\item school \\
The name of the school where a thesis was written.
\item series\\
The name of a series or set of books. When citing an entire book, the the title field gives its title and an optional series field gives the name of a series or multi-volume set in which the book is published.
\item title\\
The work's title, typed as explained in the LATEX book.
\item type\\
The type of a technical report--for example, ``Research Note''.
\item volume\\
The volume of a journal or multivolume book.
\item year\\
The year of publication or, for an unpublished work, the year it was written. Generally it should consist of four numerals, such as 1984, although the standard styles can handle any year whose last four nonpunctuation characters are numerals, such as `(about 1984)'. 
\end{itemize}
\cite{bibtexHelp}


\addtocontents{toc}{\vspace{2em}}  

\lhead{\emph{Bibliography}}  
\bibliographystyle{unsrtnat}  
\bibliography{Bibliography}  

\begin{thebibliography}{46}
\providecommand{\natexlab}[1]{#1}
\providecommand{\url}[1]{\texttt{#1}}
\expandafter\ifx\csname urlstyle\endcsname\relax
  \providecommand{\doi}[1]{doi: #1}\else
  \providecommand{\doi}{doi: \begingroup \urlstyle{rm}\Url}\fi

\bibitem[Beel et~al.(2016)Beel, Gipp, Langer, and Breitinger]{Beel2014a}
Joeran Beel, Bela Gipp, Stefan Langer, and Corinna Breitinger.
\newblock Research paper recommender systems: A literature survey.
\newblock \emph{International Journal on Digital Libraries}, \penalty0
  (4):\penalty0 305–338, 2016.
\newblock \doi{10.1007/s00799-015-0156-0}.

\bibitem[Tkaczyk et~al.()Tkaczyk, Collins, Sheridan, and Beel]{outofbox}
Dominika Tkaczyk, Andrew Collins, Paraic Sheridan, and Joeran Beel.
\newblock Machine learning vs. rules and out-of-the-box vs. retrained: An
  evaluation of open-source bibliographic reference and citation parsers.
\newblock \emph{ACM Joint Conference on Digital Libraries}.

\bibitem[Tkaczyk et~al.(2015{\natexlab{a}})Tkaczyk, Szostek, Fedoryszak,
  Dendek, and Bolikowski]{DBLP:journals/ijdar/TkaczykSFDB15}
Dominika Tkaczyk, Pawel Szostek, Mateusz Fedoryszak, Piotr~Jan Dendek, and
  Lukasz Bolikowski.
\newblock {CERMINE:} automatic extraction of structured metadata from
  scientific literature.
\newblock \emph{{IJDAR}}, 18\penalty0 (4):\penalty0 317--335,
  2015{\natexlab{a}}.
\newblock \doi{10.1007/s10032-015-0249-8}.
\newblock URL \url{https://doi.org/10.1007/s10032-015-0249-8}.

\bibitem[Tkaczyk et~al.(2014{\natexlab{a}})Tkaczyk, Szostek, Dendek,
  Fedoryszak, and Bolikowski]{DBLP:conf/das/TkaczykSDFB14}
Dominika Tkaczyk, Pawel Szostek, Piotr~Jan Dendek, Mateusz Fedoryszak, and
  Lukasz Bolikowski.
\newblock {CERMINE} - automatic extraction of metadata and references from
  scientific literature.
\newblock In Jean{-}Yves Ramel, Marcus Liwicki, Jean{-}Marc Ogier, Koichi Kise,
  and Ray Smith, editors, \emph{11th {IAPR} International Workshop on Document
  Analysis Systems, {DAS} 2014, Tours, France, April 7-10, 2014}, pages
  217--221. {IEEE} Computer Society, 2014{\natexlab{a}}.
\newblock ISBN 978-1-4799-3244-3.
\newblock \doi{10.1109/DAS.2014.63}.
\newblock URL \url{https://doi.org/10.1109/DAS.2014.63}.

\bibitem[Tkaczyk and Bolikowski(2015)]{DBLP:conf/esws/TkaczykB15}
Dominika Tkaczyk and Lukasz Bolikowski.
\newblock Extracting contextual information from scientific literature using
  {CERMINE} system.
\newblock In Fabien Gandon, Elena Cabrio, Milan Stankovic, and Antoine
  Zimmermann, editors, \emph{Semantic Web Evaluation Challenges - Second
  SemWebEval Challenge at {ESWC} 2015, Portoro{\v{z}}, Slovenia, May 31 - June
  4, 2015, Revised Selected Papers}, volume 548 of \emph{Communications in
  Computer and Information Science}, pages 93--104. Springer, 2015.
\newblock ISBN 978-3-319-25517-0.
\newblock \doi{10.1007/978-3-319-25518-7_8}.
\newblock URL \url{https://doi.org/10.1007/978-3-319-25518-7_8}.

\bibitem[Tkaczyk et~al.(2014{\natexlab{b}})Tkaczyk, Szostek, and
  Bolikowski]{tkaczyk2014grotoap2}
Dominika Tkaczyk, Pawel Szostek, and {\L}ukasz Bolikowski.
\newblock Grotoap2—the methodology of creating a large ground truth dataset
  of scientific articles.
\newblock \emph{D-Lib Magazine}, 20\penalty0 (11/12), 2014{\natexlab{b}}.

\bibitem[Giles et~al.(1998)Giles, Bollacker, and Lawrence]{Giles98citeseer}
C.~Lee Giles, Kurt~D. Bollacker, and Steve Lawrence.
\newblock Citeseer: an automatic citation indexing system.
\newblock In \emph{INTERNATIONAL CONFERENCE ON DIGITAL LIBRARIES}, pages
  89--98. ACM Press, 1998.

\bibitem[McCallum et~al.(2000)McCallum, Nigam, Rennie, and
  Seymore]{mccallum2000automating}
Andrew~Kachites McCallum, Kamal Nigam, Jason Rennie, and Kristie Seymore.
\newblock Automating the construction of internet portals with machine
  learning.
\newblock \emph{Information Retrieval}, 3\penalty0 (2):\penalty0 127--163,
  2000.

\bibitem[pub()]{pubmed}
Pubmed.
\newblock URL \url{http://www.ncbi.nlm.nih.gov/pubmed}.

\bibitem[Councill et~al.(2008)Councill, Giles, and Kan]{councill2008parscit}
Isaac~G Councill, C~Lee Giles, and Min-Yen Kan.
\newblock Parscit: an open-source crf reference string parsing package.
\newblock In \emph{LREC}, volume~8, pages 661--667, 2008.

\bibitem[Fedoryszak et~al.(2013{\natexlab{a}})Fedoryszak, Tkaczyk, and
  Bolikowski]{DBLP:conf/ercimdl/FedoryszakTB13}
Mateusz Fedoryszak, Dominika Tkaczyk, and Lukasz Bolikowski.
\newblock Large scale citation matching using apache hadoop.
\newblock In Trond Aalberg, Christos Papatheodorou, Milena Dobreva, Giannis
  Tsakonas, and Charles~J. Farrugia, editors, \emph{Research and Advanced
  Technology for Digital Libraries - International Conference on Theory and
  Practice of Digital Libraries, {TPDL} 2013, Valletta, Malta, September 22-26,
  2013. Proceedings}, volume 8092 of \emph{Lecture Notes in Computer Science},
  pages 362--365. Springer, 2013{\natexlab{a}}.
\newblock ISBN 978-3-642-40500-6.
\newblock \doi{10.1007/978-3-642-40501-3_37}.
\newblock URL \url{https://doi.org/10.1007/978-3-642-40501-3_37}.

\bibitem[Fedoryszak et~al.(2013{\natexlab{b}})Fedoryszak, Bolikowski, Tkaczyk,
  and Wojciechowski]{DBLP:series/sci/FedoryszakBTW13}
Mateusz Fedoryszak, Lukasz Bolikowski, Dominika Tkaczyk, and Krzysztof
  Wojciechowski.
\newblock Methodology for evaluating citation parsing and matching.
\newblock In Robert Bembenik, Lukasz Skonieczny, Henryk Rybinski, Marzena
  Kryszkiewicz, and Marek Niezgodka, editors, \emph{Intelligent Tools for
  Building a Scientific Information Platform - Advanced Architectures and
  Solutions}, volume 467 of \emph{Studies in Computational Intelligence}, pages
  145--154. Springer, 2013{\natexlab{b}}.
\newblock ISBN 978-3-642-35646-9.
\newblock \doi{10.1007/978-3-642-35647-6_11}.
\newblock URL \url{https://doi.org/10.1007/978-3-642-35647-6_11}.

\bibitem[{B}eel et~al.(2011){B}eel, {G}ipp, {L}anger, {G}enzmehr, {W}ilde,
  {N}ürnberger, and {P}itman]{Beel2011b}
{J}oeran {B}eel, {B}ela {G}ipp, {S}tefan {L}anger, {M}arcel {G}enzmehr, {E}rik
  {W}ilde, {A}ndreas {N}ürnberger, and {J}im {P}itman.
\newblock {I}ntroducing {M}r. {DL}ib, a {M}achine-readable {D}igital {L}ibrary.
\newblock In \emph{{P}roceedings of the 11th {ACM}/{IEEE} {J}oint {C}onference
  on {D}igital {L}ibraries ({JCDL}`11)}, pages 463--464. ACM, 2011.
\newblock \doi{10.1145/1998076.1998187}.

\bibitem[Beel et~al.(2017)Beel, Aizawa, Breitinger, and Gipp]{Beel2017g}
Joeran Beel, Akiko Aizawa, Corinna Breitinger, and Bela Gipp.
\newblock Mr. dlib: Recommendations-as-a-service (raas) for academia.
\newblock In \emph{Proceedings of the ACM/IEEE-CS Joint Conference on Digital
  Libraries (JCDL)}, pages 1--2, 2017.
\newblock \doi{10.1109/JCDL.2017.7991606}.

\bibitem[Zhang et~al.(2011{\natexlab{a}})Zhang, Zou, Le, and
  Thoma]{zhang2011structural}
Xiaoli Zhang, Jie Zou, Daniel~X Le, and George~R Thoma.
\newblock A structural svm approach for reference parsing.
\newblock \emph{BMC bioinformatics}, 12\penalty0 (3):\penalty0 S7,
  2011{\natexlab{a}}.

\bibitem[Wu et~al.(2014)Wu, Williams, Chen, Khabsa, Caragea, Ororbia, Jordan,
  and Giles]{IAAI148607}
Jian Wu, Kyle Williams, Hung-Hsuan Chen, Madian Khabsa, Cornelia Caragea,
  Alexander Ororbia, Douglas Jordan, and C.~Giles.
\newblock Citeseerx: Ai in a digital library search engine, 2014.
\newblock URL
  \url{https://www.aaai.org/ocs/index.php/IAAI/IAAI14/paper/view/8607}.

\bibitem[Lipinski et~al.(2013)Lipinski, Yao, Breitinger, Beel, and
  Gipp]{Lipinski:2013:EHM:2467696.2467753}
Mario Lipinski, Kevin Yao, Corinna Breitinger, Joeran Beel, and Bela Gipp.
\newblock Evaluation of header metadata extraction approaches and tools for
  scientific pdf documents.
\newblock In \emph{Proceedings of the 13th ACM/IEEE-CS Joint Conference on
  Digital Libraries}, JCDL '13, pages 385--386, New York, NY, USA, 2013. ACM.
\newblock ISBN 978-1-4503-2077-1.
\newblock \doi{10.1145/2467696.2467753}.
\newblock URL \url{http://doi.acm.org/10.1145/2467696.2467753}.

\bibitem[Yu and Fan(2007)]{Yu2007}
Jiangde Yu and Xiaozhong Fan.
\newblock Metadata extraction from chinese research papers based on conditional
  random fields.
\newblock In \emph{Fourth International Conference on Fuzzy Systems and
  Knowledge Discovery ({FSKD} 2007)}. {IEEE}, 2007.
\newblock \doi{10.1109/fskd.2007.394}.
\newblock URL \url{https://doi.org/10.1109/fskd.2007.394}.

\bibitem[Zhang et~al.(2011{\natexlab{b}})Zhang, Cao, and Yu]{Zhang2011}
Qing Zhang, Yong-Gang Cao, and Hong Yu.
\newblock Parsing citations in biomedical articles using conditional random
  fields.
\newblock \emph{Computers in Biology and Medicine}, 41\penalty0 (4):\penalty0
  190--194, apr 2011{\natexlab{b}}.
\newblock \doi{10.1016/j.compbiomed.2011.02.005}.
\newblock URL \url{https://doi.org/10.1016/j.compbiomed.2011.02.005}.

\bibitem[Lin et~al.(2010)Lin, Ng, Pradhan, Shah, Pietrobon, and
  Kan]{Lin:2010:EFF:1867735.1867749}
Sein Lin, Jun-Ping Ng, Shreyasee Pradhan, Jatin Shah, Ricardo Pietrobon, and
  Min-Yen Kan.
\newblock Extracting formulaic and free text clinical research articles
  metadata using conditional random fields.
\newblock In \emph{Proceedings of the NAACL HLT 2010 Second Louhi Workshop on
  Text and Data Mining of Health Documents}, Louhi '10, pages 90--95,
  Stroudsburg, PA, USA, 2010. Association for Computational Linguistics.
\newblock URL \url{http://dl.acm.org/citation.cfm?id=1867735.1867749}.

\bibitem[Pinto et~al.(2003)Pinto, McCallum, Wei, and Croft]{pinto2003table}
David Pinto, Andrew McCallum, Xing Wei, and W~Bruce Croft.
\newblock Table extraction using conditional random fields.
\newblock In \emph{Proceedings of the 26th annual international ACM SIGIR
  conference on Research and development in informaion retrieval}, pages
  235--242. ACM, 2003.

\bibitem[Peng and McCallum()]{pengaccurate}
F~Peng and A~McCallum.
\newblock Accurate information extraction from research papers using
  conditional random fields. retrieved on april 13, 2013.

\bibitem[Groza et~al.(2012)Groza, Grimnes, and Handschuh]{groza2012reference}
Tudor Groza, Gunnar~AAstrand Grimnes, and Siegfried Handschuh.
\newblock Reference information extraction and processing using conditional
  random fields.
\newblock \emph{Information Technology and Libraries (Online)}, 31\penalty0
  (2):\penalty0 6, 2012.

\bibitem[Kr{\"a}mer()]{michel}
Michel Kr{\"a}mer.
\newblock citeproc-java: A citation style language (csl) processor for java.
\newblock URL \url{http://michel-kraemer.github.io/citeproc-java/}.

\bibitem[Gao et~al.(2012)Gao, Qi, Tang, Lin, and Liu]{gao2012web}
Liangcai Gao, Xixi Qi, Zhi Tang, Xiaofan Lin, and Ying Liu.
\newblock Web-based citation parsing, correction and augmentation.
\newblock In \emph{Proceedings of the 12th ACM/IEEE-CS joint conference on
  Digital Libraries}, pages 295--304. ACM, 2012.

\bibitem[Cameron(1997)]{cameron1997universal}
Robert~D Cameron.
\newblock A universal citation database as a catalyst for reform in scholarly
  communication.
\newblock \emph{First Monday}, 2\penalty0 (4), 1997.

\bibitem[Lawrence et~al.(1999)Lawrence, Giles, and
  Bollacker]{lawrence1999digital}
Steve Lawrence, C~Lee Giles, and Kurt Bollacker.
\newblock Digital libraries and autonomous citation indexing.
\newblock \emph{Computer}, 32\penalty0 (6):\penalty0 67--71, 1999.

\bibitem[Cortez et~al.(2007)Cortez, da~Silva, Gon{\c{c}}alves, Mesquita, and
  de~Moura]{cortez2007flux}
Eli Cortez, Altigran~S da~Silva, Marcos~Andr{\'e} Gon{\c{c}}alves, Filipe
  Mesquita, and Edleno~S de~Moura.
\newblock Flux-cim: flexible unsupervised extraction of citation metadata.
\newblock In \emph{Proceedings of the 7th ACM/IEEE-CS joint conference on
  Digital libraries}, pages 215--224. ACM, 2007.

\bibitem[Jewell(2000)]{jewell2000paracite}
Michael Jewell.
\newblock Paracite: An overview, 2000.

\bibitem[Yin et~al.(2004)Yin, Zhang, Deng, and Yang]{yin2004metadata}
Ping Yin, Ming Zhang, ZhiHong Deng, and DongQing Yang.
\newblock Metadata extraction from bibliographies using bigram hmm.
\newblock In \emph{International Conference on Asian Digital Libraries}, pages
  310--319. Springer, 2004.

\bibitem[Hetzner(2008)]{hetzner2008simple}
Erik Hetzner.
\newblock A simple method for citation metadata extraction using hidden markov
  models.
\newblock In \emph{Proceedings of the 8th ACM/IEEE-CS joint conference on
  Digital libraries}, pages 280--284. ACM, 2008.

\bibitem[Lafferty et~al.(2001)Lafferty, McCallum, and
  Pereira]{lafferty2001conditional}
John Lafferty, Andrew McCallum, and Fernando~CN Pereira.
\newblock Conditional random fields: Probabilistic models for segmenting and
  labeling sequence data.
\newblock 2001.

\bibitem[Bradham()]{bradham}
Elyssa Bradham.
\newblock \emph{Hidden Markov Model and some applications in handwriting
  recognition Published byElyssa Bradham Modified over 3 years ago 52 Embed
  Download presentation Presentation on theme: "Hidden Markov Model and some
  applications in handwriting recognition"— Presentation transcript: 1 Hidden
  Markov Model and some applications in handwriting recognition}.
\newblock URL \url{http://slideplayer.com/slide/1416164/4/images/17/Hidden
  Markov Model Start Rainy Sunny Walk Shop Clean Example (cont):.jpg}.

\bibitem[Lopez and Romary(2015)]{DBLP:journals/ercim/LopezR15}
Patrice Lopez and Laurent Romary.
\newblock {GROBID} - information extraction from scientific publications.
\newblock \emph{{ERCIM} News}, 2015\penalty0 (100), 2015.
\newblock URL
  \url{http://ercim-news.ercim.eu/en100/r-i/grobid-information-extraction-from-scientific-publications}.

\bibitem[Cuong et~al.(2015)Cuong, Chandrasekaran, Kan, and
  Lee]{DBLP:conf/jcdl/CuongCKL15}
Nguyen~Viet Cuong, Muthu~Kumar Chandrasekaran, Min{-}Yen Kan, and Wee~Sun Lee.
\newblock Scholarly document information extraction using extensible features
  for efficient higher order semi-crfs.
\newblock In Paul Logasa~Bogen II, Suzie Allard, Holly Mercer, Micah Beck,
  Sally~Jo Cunningham, Dion~Hoe{-}Lian Goh, and Geneva Henry, editors,
  \emph{Proceedings of the 15th {ACM/IEEE-CE} Joint Conference on Digital
  Libraries, Knoxville, TN, USA, June 21-25, 2015}, pages 61--64. {ACM}, 2015.
\newblock ISBN 978-1-4503-3594-2.
\newblock \doi{10.1145/2756406.2756946}.
\newblock URL \url{http://doi.acm.org/10.1145/2756406.2756946}.

\bibitem[Tkaczyk et~al.(2015{\natexlab{b}})Tkaczyk, Tarnawski, and
  Bolikowski]{DBLP:journals/dlib/TkaczykTB15}
Dominika Tkaczyk, Bartosz Tarnawski, and Lukasz Bolikowski.
\newblock Structured affiliations extraction from scientific literature.
\newblock \emph{D-Lib Magazine}, 21\penalty0 (11/12), 2015{\natexlab{b}}.
\newblock \doi{10.1045/november2015-tkaczyk}.
\newblock URL \url{https://doi.org/10.1045/november2015-tkaczyk}.

\bibitem[Tkaczyk et~al.(2014{\natexlab{c}})Tkaczyk, Szostek, Dendek,
  Fedoryszak, and Bolikowski]{tkaczyk2014cermine}
Dominika Tkaczyk, Pawel Szostek, Piotr~Jan Dendek, Mateusz Fedoryszak, and
  Lukasz Bolikowski.
\newblock Cermine--automatic extraction of metadata and references from
  scientific literature.
\newblock In \emph{Document Analysis Systems (DAS), 2014 11th IAPR
  International Workshop on}, pages 217--221. IEEE, 2014{\natexlab{c}}.

\bibitem[Halevy et~al.(2009)Halevy, Norvig, and
  Pereira]{halevy2009unreasonable}
Alon Halevy, Peter Norvig, and Fernando Pereira.
\newblock The unreasonable effectiveness of data.
\newblock \emph{IEEE Intelligent Systems}, 24\penalty0 (2):\penalty0 8--12,
  2009.

\bibitem[Seymore et~al.(1999)Seymore, McCallum, and
  Rosenfeld]{seymore1999learning}
Kristie Seymore, Andrew McCallum, and Roni Rosenfeld.
\newblock Learning hidden markov model structure for information extraction.
\newblock In \emph{AAAI-99 workshop on machine learning for information
  extraction}, pages 37--42, 1999.

\bibitem[{J}oeran Beel and {B}ela Gipp(2010)]{Beel10c}
{J}oeran Beel and {B}ela Gipp.
\newblock On the robustness of google scholar against spam.
\newblock In \emph{{P}roceedings of the 21st {ACM} {C}onference on {H}ypertext
  and {H}ypermedia ({HT}'10)}, pages 297--298, Toronto (CA), June 2010. {ACM}.
\newblock \doi{10.1145/1810617.1810683}.

\bibitem[Beel and Gipp(2010)]{Beel2010}
Joeran Beel and Bela Gipp.
\newblock Academic search engine spam and google scholar's resilience against
  it.
\newblock \emph{Journal of Electronic Publishing}, 13\penalty0 (3), \#dec\#
  2010.
\newblock \doi{10.3998/3336451.0013.305}.

\bibitem[sea(2017)]{searchnewcentral_2017}
How to: Scrape search engines without pissing them off, Feb 2017.
\newblock URL
  \url{https://searchnewscentral.com/blog/2011/09/28/how-to-scrape-se
  arch-engines-without-pissing-them-off/}.

\bibitem[CSL()]{CSL}
Citation style language.
\newblock URL \url{http://citationstyles.org/}.

\bibitem[Lopez(2009)]{lopez2009grobid}
Patrice Lopez.
\newblock Grobid: Combining automatic bibliographic data recognition and term
  extraction for scholarship publications.
\newblock In \emph{International Conference on Theory and Practice of Digital
  Libraries}, pages 473--474. Springer, 2009.

\bibitem[Wei et~al.(2016)Wei, Wu, Zhao, Deng, Ersotelos, Parvinzamir, Liu, Liu,
  and Dong]{unknown}
Hui Wei, Shaopeng Wu, Youbing Zhao, Zhikun Deng, Nikolaos Ersotelos, Farzad
  Parvinzamir, Baoquan Liu, Enjie Liu, and Feng Dong.
\newblock Data mining, management and visualization in large scientific
  corpuses, 04 2016.

\bibitem[bib()]{bibtexHelp}
Bibtex entry types, field types and usage hints.
\newblock URL \url{https://www.openoffice.org/bibliographic/bibtex-defs.html}.

\end{thebibliography}
\bibliographystyle{IEEEtran}

\end{document}